\documentclass[prx,longbibliography,nofootinbib,twocolumn,showkeys,superscriptaddress,preprintnumbers,floatfix]{revtex4-1}

\usepackage{etex}
\usepackage{subfigure}
\usepackage{ifpdf}
\usepackage{hyperref}
\usepackage{dcolumn}
\usepackage{url}
\usepackage{amsmath}
\usepackage{amscd}
\usepackage{amsfonts}
\usepackage{amssymb}
\usepackage{bm}   
\usepackage{bbm}
\usepackage{verbatim}
\usepackage{comment}
\usepackage{stmaryrd}
\usepackage{amsthm}
\usepackage{xcolor}
\usepackage{setspace}
\usepackage{braket}
\usepackage{empheq}
\usepackage{perpage}
\usepackage[english]{babel}
\MakePerPage{footnote}

\usepackage{tikz}
\usetikzlibrary{arrows}
\usetikzlibrary{automata}
\usetikzlibrary{decorations.pathreplacing}
\usetikzlibrary{positioning}
\usetikzlibrary{plotmarks}
\usetikzlibrary{calc}
\usetikzlibrary{patterns}
\usetikzlibrary{external}
\usepackage{pgfplots}
\pgfplotsset{compat=newest}

\usepackage{graphicx}
\graphicspath{ {./Diagrams/} }

\theoremstyle{plain}    
\theoremstyle{plain}    
\theoremstyle{plain}    
\theoremstyle{plain}    
\theoremstyle{plain}    
\theoremstyle{plain}    
\theoremstyle{plain}    
\theoremstyle{plain}    
\theoremstyle{plain}    
\theoremstyle{plain}    
\theoremstyle{plain}    
\theoremstyle{plain}

\newcommand{\CausalState}   { \mathcal{S} }

\newcommand{\forward}{+}
\newcommand{\reverse}{-}
\newcommand{\forwardreverse}{\pm} 

\newcommand{\FutureCausalState} { {\CausalState}^{\forward} }

\newcommand{\PastCausalState}   { {\CausalState}^{\reverse} }

\newcommand{\lastindex}[2]{
  \edef\tempa{0}
  \edef\tempb{#2}
  \ifx\tempa\tempb
    \edef\tempc{#1}
  \else
    \edef\tempa{0}
    \edef\tempb{#1}
    \ifx\tempa\tempb
      \edef\tempc{#2}
    \else
      \edef\tempc{#1+#2}
    \fi
  \fi
  \tempc
}

\newcommand{\CSjoint}[1][,]{
   \edef\tempa{:}
   \edef\tempb{#1}
   \ifx\tempa\tempb
      \ensuremath{\FutureCausalState\!#1\PastCausalState}
   \else
      \ensuremath{\FutureCausalState#1\PastCausalState}
   \fi
}

\newif\ifpm
\edef\tempa{\forwardreverse}
\edef\tempb{\pm}
\ifx\tempa\tempb
   \pmfalse
\else
   \pmtrue
\fi

\newcommand{\R}[2]{R_{#1\rightarrow#2}}

\usepackage{verbatim}

\begin{document}


\title{Large Dense Thermodynamic Networks Nearly Minimize Entropy Production}

\author{Kyle J. Ray}
\email{kylejray@gmail.com}
\affiliation{Complexity Sciences Center and Department of Physics and Astronomy,
University of California at Davis, One Shields Avenue, Davis, CA 95616}

\author{Alexander B. Boyd}
\email{alecboy@gmail.com}
\affiliation{Beyond Institute for Theoretical Science, San Francisco, CA}
\affiliation{University of Sussex, Falmer, Brighton, UK BN1 9RH}
\affiliation{Corresponding Author}

\begin{abstract}
Many have speculated whether nonequilibrium systems obey principles of maximum or minimum entropy production. In this work, we use stochastic thermodynamics to derive the condition for the minimum entropy production state (MEPS) for continuous-time Markov chains (CTMCs), even far from equilibrium. We show that real nonequilibrium steady states (NESS) generally violate both the Minimum and Maximum Entropy Production (MINEP and MAXEP) principles. However, through numerical sampling of large densely-connected ($N_\text{edges}= \mathcal{O}(N_\text{states}^2)$) CTMCs, we find that as system size increases, the steady-state entropy production tends to converge to a value close to the minimum.  By contrast, sparsely-connected ($N_\text{edges}= \mathcal{O}(N_\text{states})$) thermodynamic networks become less efficient. This suggests that large dense nonequilibrium networks may self-organize to use thermodynamic resources efficiently, offering a nuanced perspective on the longstanding debate between MAXEP and MINEP.
\end{abstract}

\maketitle

\section{Introduction}

``Nature abhors a gradient.'' -Eric Schneider \cite{schneider2005into}

Particles diffuse toward lower pressure, heat seeps from hot to cold, and electrons drift towards lower electric potential--all manifestations of the Second Law of Thermodynamics' dictate that entropy is non-decreasing, forever drawing us towards the ultimate equilibrium and heat death of the universe.  However, the question remains: what path do we take to this end? Does the universe slowly meander towards high entropy, like a river snaking through a valley, or does it crash towards equilibrium with the power of a waterfall?  Is nonequilibrium truly ``abhorrent'' to Nature, demanding rapid dissipation, or is it merely ``disagreeable,'' allowing for more gradual scenic paths to the maximum entropy (MAXENT) state \cite{jaynes1980minimum}.

Using the tools of modern stochastic thermodynamics \cite{seifert2012stochastic}, we address the question of whether the rate of dissipation of thermodynamic resources is naturally maximized, minimized, or neither.  Thermodynamic resources, from chemical potential to heat, ultimately contribute to the \emph{entropy production}, which serves as the ultimate measure of irreversibility in physics \cite{seif2021machine,batalhao2015irreversibility, landi2021irreversible}. Past analyses have attempted to answer this question by arguing for principles of Maximum Entropy Production (MAXEP)\cite{dewar2003information,dewar20064,endres2017entropy} or Minimum Entropy Production (MINEP) \cite{onsager1931reciprocal, prigogine1978time}.  The search for physical laws of emergent organization in nonequilibrium systems \cite{rupe2024principles} to counterpose the unyielding Second Law of thermodynamics is ongoing.  Amidst the spontaneous decay of structure through entropy production, we also witness emergence of complexity from simple chemical patterns \cite{zhabotinsky1991history} to life \cite{michaelian2022non}, motivating this search.  Despite in-depth investigations from researchers across disciplines, there is not yet a consensus on the universal behavior of entropy production, perhaps best illustrated by the fact that advocates of both MAXEP and MINEP still coexist in the scientific conversation. Translating into familiar (but often conflated) acronyms: this article examines whether nonequilibrium dynamics more closely resemble MAXEP, MINEP, or neither in the pursuit of the MAXENT equilibrium state?

Perhaps the core challenge for establishing universal principles of organization and entropy production is discovering a framework that encompasses a sufficiently broad class of nonequilibrium systems.  For instance, Prigogine and Onsager were phenomenologically motivated by thermodynamic systems; they considered systems in the linear-response regime, despite knowing that nonlinearity is a typical phenomenon in complex systems \cite{onsager1931reciprocal,Prig68a}.  The search for principles of organization for nonequilibrium systems is more than a century old, extending back to the Helmholtz minimum dissipation theorem for fluid flow published in 1868 \cite{guazzelli2011physical}, but it is only in the last two decades that stochastic thermodynamics, built from first principles fluctuation theorems, has established general relationship between stochastic dynamics and entropy production \cite{Croo99a,Jarz00,roldan2010estimating}.  This relationship has been validated at a variety of experimental scales in quantum heat exchange \cite{micadei2021experimental}, computing nanodevices \cite{wimsatt2021harnessing,Jun14a}, Maxwellian demons \cite{saha2023information,tang2025nonequilibrium}, biological systems \cite{hayashi2010fluctuation,martinez2019inferring,gnesotto2018broken}, and even macroscopic phenomenon \cite{skinner2021estimating}.  Stochastic thermodynamics presents powerful tools to analyze entropy production for virtually any physical system.  With this framework, we can derive the Minimum Entropy Production State  (MEPS) distribution \cite{riechers2024thermodynamically} directly from the underlying system dynamics and compare it quantitatively with the Nonequilibrium Steady State (NESS) distribution of the corresponding stochastic process.  Both the NESS and MEPS describe \emph{probability distributions over system states} rather than specific configurations.  The MEPS can be interpreted as the prior distribution of the thermodynamic computation, i.e. the distribution over which the system performs its physical information processing most efficiently \cite{kolchinsky2017dependence}.  The resulting picture is more nuanced than the classical formulations, but surprisingly echoes insights originally articulated by Prigogine and other pioneers of nonequilibrium thermodynamics.

We specialize our analysis to the dynamics of linear rate equations, otherwise known as Continuous Time Markov Chains (CTMCs) or master equations, for which the continuous probabilistic dynamics are directly connected to the consumption of thermodynamic resources \cite{wolpert2019stochastic,van2023thermodynamic}.  Within this framework, we derive continuous dynamics that arrive at the MEPS for any such system, and show that the NESS and MEPS are only equivalent in the case of entropy-production symmetry.  

While the application of this framework can be more general, we analyze discrete energy levels whose dynamics are driven out of equilibrium by external pumping \cite{brown2017allocating}.  We begin with a simple 3-state system with a tunable pump that drives cyclical transitions, much like a laser \cite{andrews2010off}.  The resulting dynamics are governed by Arrhenius rate equations \cite{busiello2021dissipation}. We see a variety of entropic behaviours, depending on the pumping parameter. In general, the steady state of the system does not achieve minimum entropy production and never achieves the maximum entropy production.  However, interestingly, there are regimes in which the NESS produces significant entropy, but nearly achieves the minimum regardless. In these regimes, we can say that the system spontaneously organizes to produce much less entropy than a uniformly distributed (unbiased) configuration of the system.  

We further probe the typicality of minimum entropy production by considering randomly generated dense thermodynamic networks.  
We compare the steady state of the resulting Arrhenius rate equations to the minimum entropy production distribution.  Again, we find that entropy production is not generally minimized.  However, we find that as the number of system states increases, the steady-state entropy production converges on a value that is very close to the minimum entropy production. While our small toy model can be shown to easily violate the MINEP principle, instances of strong MINEP violation diminish as the system size increases.

To put the results of dense thermodynamic networks into relief, we consider counterexamples where MINEP is strongly violated. We introduce state-exchange asymmetry into our networks through sparsity and time-scale separation; in both cases, the asymmetry leads to violations of entropy production-symmetry, yielding thermodynamically inefficient steady states.  This study suggests that there is a kernel of truth in the ideas of MINEP, that, despite counterexamples, many thermodynamic systems spontaneously organize to make the best use of thermodynamic resources \cite{boyd2022thermodynamic, boyd2024thermodynamic, gold2019self}.

\section{Rate Equation Thermodynamics}

Many stochastic thermodynamic systems can be described through rate equations that implement the continuous time Markovian evolution of a density $p$ over states $\mathcal{S}$.  The rate of change of the probability $p(s')$ of any state $s'$ is linearly related to the current probability distribution
\begin{align}
    \partial_t p(s') = \sum_{s} p(s)R_{s \rightarrow s'}.
\end{align}
The term $R_{s \rightarrow s'}$ is the rate of transitioning from state $s$ to state $s'$.  These equations can also be written in matrix-vector form $\partial_t p =Rp$.  These describe a broad class of thermodynamic processes, including nonequilibrium systems in which the steady state $\pi$ that satisfies $R\pi=0$ produces entropy.

For time-reversal invariant states $s=s^\dagger$, the relative rate of forward and reverse transitions determines the amount of environmental entropy production (measured in units of $k_B$) associated with transitions between the two states
\begin{align}
\label{eq:EntFlow}
\Delta S^\text{env}_{s \rightarrow s'}= \ln \frac{R_{s \rightarrow s'}}{R_{s' \rightarrow s}}.
\end{align}
The corresponding entropy production rate (EPR) is the average environmental entropy production plus the rate of change in the surprisal of the system
\begin{align}
\label{Eq:EntProd}
\sigma(R,p) &= \sum_{s,s'}p(s) R_{s \rightarrow s'} \Delta S^\text{env}_{s \rightarrow s'}-\sum_{s'}(\partial_tp(s')) \ln p(s') \nonumber \\
&= \sum_{s,s'}p(s) R_{s \rightarrow s'}\ln\frac{p(s) R_{s \rightarrow s'}}{p(s') R_{s' \rightarrow s}}.
\end{align}
Specifically, this is the entropy production of a continuous-time Markov chain (CTMC) \cite{wolpert2019stochastic}, where the associated states are time-reversal symmetric. One might interpret the claims of Prigogine to mean that entropy production is minimized in steady state.  In the stochastic thermodynamic context, the NESS distribution $\pi$ produces entropy that we can calculate directly
\begin{align}
    \sigma_\text{NESS} \equiv \sigma(R,\pi).
\end{align}

Using linear rate equation dynamics to analyze nonequilibrium behavior is a well-established tradition, spanning over decades \cite{RevModPhys.48.571}. Despite its apparent simplicity, it remains the framework for a large fraction of modern stochastic thermodynamics literature, from TURs and speed limits \cite{barato2015thermodynamic, shiraishi2018speed} to biochemical computing \cite{gunawardena2012linear, cabello2025information, floyd2024limits}. Here, analyses of non-Markovian behaviour \cite{ray2023thermodynamic}, nonlinear rate equations \cite{dal2023geometry},  and time-antisymmetric variables \cite{boyd2021time, boyd2025time} are exceptions that prove the rule.  Within this broad class of processes, we investigate the extremality of entropy production in comparison with steady state behavior.  
\section{Maximum Entropy Production}

We analyze the claim that systems self-organize into the MAXEP distribution by considering the distribution $M_R$ that maximizes the entropy production for a dynamical system with rate equation $R$:
\begin{align}
    M_R \equiv \underset{p}{\text{argmax }} \sigma(R,p).
\end{align}
For the sake of simplicity, let us assume that the entropy flow into the environment $\Delta S^\text{env}_{s \rightarrow s'} $ is finite for any transition $s \rightarrow s'$.  We can look at the first line of Eq. \ref{Eq:EntProd}
to identify a simple way to maximize the entropy production.  If we choose a state $s_0$ to have zero probability $p(s_0)=0$, then, as long as there is nonzero probability flow to that state, its probability will increase with time $\partial_t p(s_0)$.  Thus, in the change in system entropy, the term $-\partial_t p(s_0) \ln p(s_0)$ contributes positive infinity to the overall entropy production rate.  Because all other terms with nonzero probability are finite, the overall entropy production is infinite and therefore maximized.  This affirms the result that MAXEP distributions exist on the edge of the probability simplex \cite{riechers2024thermodynamically}.

The MAXEP distributions are those for which a number of states have zero probability.  The system entropy rises infinitely fast as probability mass rushes into these unoccupied states instantaneously.  These states are not typical steady-states $\pi$ that satisfy $\partial_t \pi =0$.  Steady states are certain not to yield infinite entropy production in this way, because their system entropy must be fixed in time.  For this reason, in the context of thermodynamic continuous-time Markov chains, the notion of MAXEP presented above does not hold.

\section{Minimum Entropy Production State}
\label{sec:MEPS_derivation}
In analyzing the claim of MINEP, the countervailing perspective to MAXEP, we wish to determine the minimum entropy state (MEPS) for system dynamics that are determined by rate equation $R$.  Note that, like the NESS, the MEPS is not a specific state configuration $s$ of the system, but a distribution over those states.  We denote the MEPS distribution for a rate equation $R$ by $m_R$:
\begin{align}
    m_R \equiv \underset{p}{\text{argmin }} \sigma(R,p).
\end{align}
As discussed in the previous section, the maximum entropy state produces infinite entropy, so we may apply the method of Lagrange multipliers to the entropy production to find the minimum.  We set the partial derivative with respect to the probability
\begin{align}
    \partial_{m_R(s)}\sigma (R,m_R) = \lambda \partial_{m_R(s)} g(m_R),
\end{align}
where $g(p)=\sum_{s'}p(s')=1$ is the constraint that $p$ is normalized, and $\lambda$ is the Lagrange multiplier.  Solving this, we find an equation that describes the dynamics of the MEPS
\begin{align}
\label{eq:meps_dynamics}
    \partial_t \ln m_R(s) = \sigma(R,m_R,s)-\sigma(R,m_R),
\end{align}
where $\sigma(R,p,s) \equiv \sum_{s'}R_{s \rightarrow s'}\ln\frac{p(s) R_{s \rightarrow s'}}{p(s') R_{s' \rightarrow s}} $ is the contribution to the entropy production from  state $s$ ($\sigma(R,p)=\sum_s p(s) \sigma(R,p,s)$), and $\partial_t \ln m_R(s) \equiv \frac{\sum_{s'}m_R(s')R_{s' \rightarrow s}}{m_R(s)}$ is the rate of change in log-probability of state $s$ were the system prepared in the MEPS and allowed to evolve according to the rate-equation dynamics. A detailed derivation is available in App. \ref{app:Minimum Entropy Production}.  We observe a few features of the MEPS that solves this equation:
\begin{enumerate}
\item The MEPS evolves toward states that dissipate greater entropy than the average ($\sigma(R,m_R,s)-\sigma(R,m_R)>0$) and away from states that dissipate less than the average ($\sigma(R,m_R,s)-\sigma(R,m_R)<0$).  
\item Eq. \ref{eq:meps_dynamics} implies the MEPS and the stationary distribution ($\dot{p}=0$) are the same if and only if every state $s$ produces the same amount of entropy in steady state 
\begin{align}
\label{eq:Ent_Symmetry}
\pi & = m_R
\\ & \Leftrightarrow \nonumber
\\ \sigma(R,\pi,s) &=\sigma(R,\pi).\nonumber
\end{align}  
This equality provides a theoretical connection between symmetry and entropy production, because MINEP holds exclusively for systems that exhibit this special \emph{entropy-production symmetry}.
\item We can discover the MEPS by setting a dynamical equation for the probability with computational time $\tau$
\begin{align}
    \partial_\tau p(s)= \partial_t p(s) -p(s)( \sigma(R,p,s)-\sigma(R,p)).
\end{align}
When this equation of motion reaches its steady state $\partial_\tau p(s)=0$, the condition for minimum entropy production in Eq. \ref{eq:meps_dynamics} is reached.  Applying this dynamic is a strategy for discovering thermodynamically optimal states, in line with Ref. \cite{riechers2024thermodynamically}. $\tau$ here does not correspond to any physical timescale.
\end{enumerate}

The fictitious dynamics that lead to EPR minimization can also be expressed in terms of log-probabilities
\begin{align}
    \partial_\tau \ln p(s) = \partial_t \ln p(s) +\sigma(R,p)-\sigma(R,p,s).
\end{align}
This resembles the evolution of strategy space in a game, where $\sigma(R,p)-\sigma(R,p,s)$ is the payoff of state $s$ and $\partial_t p(s)$ are the inherent dynamics of the system \cite{sato2005stability}.

\section{Comparison to Traditional MINEP}

The concept of minimum entropy production has a long history \cite{jaynes1980minimum,riechers2024thermodynamically, rupe2024principles}.  It is an intriguing possibility that nonequilibrium systems naturally self-organize to make the most efficient use of free energy resources. There are hints of such a general principle as far back as Kirchhoff's law, which implies that the current in a circuit distributes itself to minimize the heat dissipation given a particular externally applied field \cite{jaynes1980minimum}.  

Prigogine and Onsager extended this idea to any linear system with fluxes (or transition rates) $\{J_z\}_{z \in \mathcal{Z}}$ and generalized forces $\{X_z \}_{z \in \mathcal{Z}}$ that drive those transitions \cite{onsager1931reciprocal, prigogine1978time}.  We may think of $J_z$ like the rate of a particular chemical reaction $z \in \mathcal{Z}$ and $X_z$ as the corresponding change in chemical potential.  We refer to these as fluxes rather than rates so as not to confuse ourselves with the rate matrix $R$.  Prigogine claims the assumption of ``local equilibrium'' leads to the corresponding rate of entropy production as the fluxes times their driving forces
\begin{align}
\sigma(J,X)= \sum_{z} J_z X_z.
\end{align}
Applying the assumption of linearity, meaning that the response $J_z$ to forces is linear, yields an expression for the flux
\begin{align}
    J_{z} = \sum_{z'}L_{zz'}X_{z'},
\end{align}
and the entropy production as a function of the applied forces
\begin{align}
    \sigma(X)= \sum_{z,z'}L_{z,z'}X_zX_{z'}.
\end{align}

Demonstrating the minimum entropy production principle requires selecting a subset of external forces $\mathcal{Z}'\subset \mathcal{Z}$ as fixed, and minimizing the entropy production with respect to the other forces $z \in \mathcal{Z}-\mathcal{Z}'$
\begin{align}
    \partial_{X_z}\sigma(J,X)=0.
\end{align}
Applying the Onsager reciprocal relation $L_{zz'} =L_{z'z}$ \cite{onsager1931reciprocal}, leads to the conclusion that this is equivalent to zero fluxes,
\begin{align}
    J_z=0,
\end{align}
which has been described as steady state \cite{jaynes1980minimum}.

To highlight both similarities and differences, we draw a comparison to stochastic thermodynamic entropy production, which is not necessarily rooted in the idea of fluxes and forces.  The expression for entropy production rate in Eq. \ref{Eq:EntProd} is general for any time-reversal symmetric states that evolve according to continuous-time Markov dynamics \cite{wolpert2019stochastic}.  This expression can be rewritten in the convenient form
\begin{align}
    \sigma(R,p) = \sum_{s,s'} (p(s) R_{s \rightarrow s'}-p(s') R_{s' \rightarrow s}) \ln (p(s) R_{s \rightarrow s'}),
\end{align}
where in place of $z$, we are summing over all transitions from $s \rightarrow s'$.  In this case, it is sensible to define the flux for transition $s \rightarrow s'$ as the rate of probability flow from $s$ to $s'$ minus the flow in the reverse direction
\begin{align}
J_{s \rightarrow s'}(R,p) \equiv p(s) R_{s \rightarrow s'}-p(s') R_{s' \rightarrow s}.
\end{align}
If we hope to reproduce the form of Prigogine's entropy production ($\sum_{s,s'}J_{s \rightarrow s'}X_{s\rightarrow s'}$), then we are left to choose the corresponding force as
\begin{align}
    X_{s \rightarrow s'}(R,p)= \ln (p(s)R_{s \rightarrow s'}).
\end{align}

With the parallel between macroscopic thermodynamic and stochastic thermodynamic entropy production made explicit, we can quickly see why the derivation of minimum entropy production fails to extend to our case:
\begin{enumerate}
\item \emph{Nonlinearity:}  With these definitions, fluxes can still be seen as direct responses to forces, but they are nonlinear
\begin{align}
    J_{s \rightarrow s'}=e^{X_{s \rightarrow s'}}-e^{X_{s' \rightarrow s}}.
\end{align}
Polettini proves that MINEP holds in the linear-response regime \cite{polettini2011macroscopic}, but this is not the most general case.
\item \emph{Constraints on Forces:} The forces $X_{s \rightarrow s'}(R,p)=\ln p(s) R_{s \rightarrow s'}$ cannot be totally free variables.  The state probability $p(s)$ is normalized, meaning that the constraint $\sum_s p(s)=1$ must be applied when using the method of Lagrange multipliers to optimize entropy production.
\item \emph{False Steady-State:} While zero flux $J_{s \rightarrow s'}=0$ may sound like a steady state, this is only true when detailed balance $p(s)R_{s \rightarrow s'}=p(s') R_{s' \rightarrow s}$ is satisfied.  If detailed balance holds in the steady-state, then the system is in thermodynamic equilibrium, which precludes nonequilibrium behavior.  A nonequilibrium system and its corresponding NESS $\pi$ will violate the condition $J_{s \rightarrow s'}=0$.
\end{enumerate}

With these differences in mind, it is reasonable to expect that MINEP is not satisfied in general for stochastic thermodynamic systems.  We confirm this intuition with numerical examples where we directly compare the minimum entropy production to that of the steady state entropy production.  Indeed, we find that there is a difference for fairly simple models.  However, we will also see intriguing examples that suggest that, while MINEP  is not strictly true in all cases, it approximates the truth for typical thermodynamic systems.

\section{Comparing NESS to MEPS}

\begin{figure}
\includegraphics[width=.6\columnwidth]{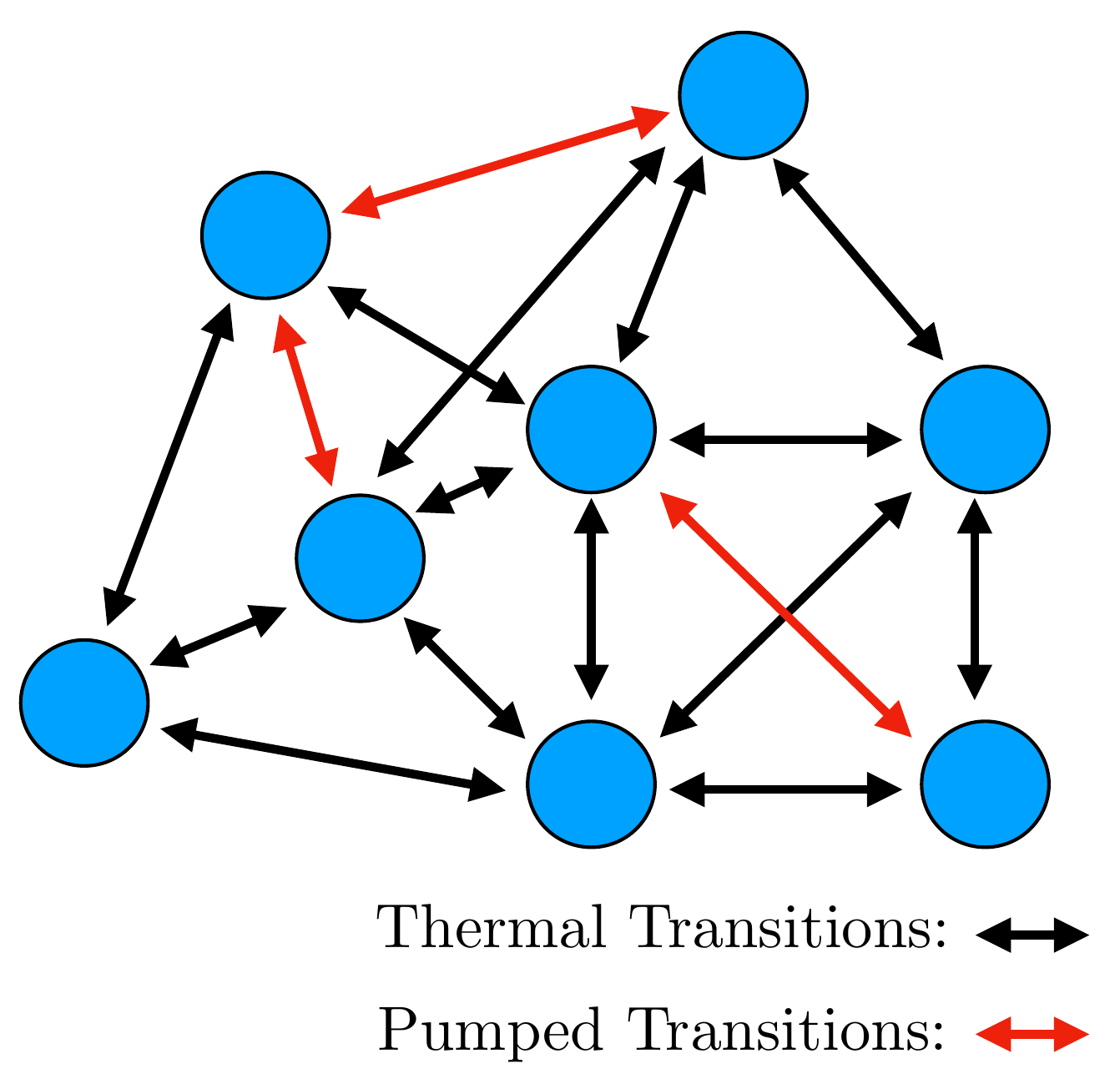}
\caption{\emph{A nonequilibrium network of thermal transitions (arrows) between states (blue circles):} The states each have an energy.  The energy difference between two states is the heat flow into the environment when a thermal transition is made between those states.  A pumped transition, by contrast, has an additional external force that contributes to the heat production and the transition rates.}
\label{fig:PumpedNetwork} 
\end{figure}

\subsection{Pumped Arrhenius Equations}

To examine the claims of MINEP, we consider a system $\mathcal{S}$ in contact with a single thermal reservoir at temperature $T$.  Given energy levels $E(s)$ and energy barriers between states $E^\text{barrier}_{s \leftrightarrow s'}=E^\text{barrier}_{s' \leftrightarrow s}$ between states, the Arrhenius equation gives rates
\begin{align}
    R_{s \rightarrow s'}=K e^{\frac{E(s)-E^\text{barrier}_{s \leftrightarrow s'}}{k_B T}}.
\end{align}

Here, the pre-exponential factor $K$ is a constant, meaning that the transition rate from $s$ to $s'$ decreases exponentially as the depth of the energy well $E^\text{barrier}_{s \leftrightarrow s'}-E(s)$ increases.  Because the environmental entropy production is proportional to heat $\Delta S^\text{env}_{s \rightarrow s'}=Q_{s \rightarrow s'}/T$ for a system in contact with a single thermal reservoir, this form of the rate equation guarantees that the heat is equal to minus the change in energy
\begin{align*}
    Q_{s \rightarrow s'}=E(s)-E(s')
\end{align*}
Moreover, the system obeys detailed balance with the steady state given by the Boltzmann distribution 
\begin{align*}
    \pi(s)= \frac{e^{-E(s)/k_B T}}{Z}.
\end{align*}  Such systems will necessarily dissipate zero entropy in steady state, and so trivially satisfy MINEP.  To make such a system nonequilibrium, we add nonequilibrium pumping.

Biochemical resources like ATP keep biological systems in a nonequilibrium steady state by pumping certain transitions that would be otherwise improbable.  We generalize this idea by adding an external pump ``force'' $F_{s \rightarrow s'}$ to certain transitions.  This is a force in that it functions like a difference in energy between two states, much like $-\partial_x V(x)$, by modifying the rate by an exponential factor in the transition rate:
\begin{align}
\label{eq:Pumps}
    R_{s \rightarrow s'}=K  e^{\frac{F_{s \rightarrow s'}+E(s)-E^\text{barrier}_{s \leftrightarrow s'}}{k_B T}}.  
\end{align} 
This echoes (and is slightly more general than) an emerging framework for biochemical and general nonequilibrium processes \cite{owen2020universal, aslyamov2024nonequilibrium}, and is sufficiently general to express any rate equation dynamics.

Regardless of its nonequilibrium driving, this system produces entropy flow according to Eq. \ref{eq:EntFlow}, as any nonequilibrium CTMC must.  Because this entropy must flow into a single heat bath, the system obeys the local detailed balance relation
\begin{align}
    Q_{s \rightarrow s'} =k_B T \ln \frac{R_{s \rightarrow s'}}{R_{s' \rightarrow s}}.
\end{align}
Thus, the additional forces result in additional heat dissipation
\begin{align}
    Q_{s \rightarrow s'}= F_{s \rightarrow s'}-F_{s' \rightarrow s}+E(s)-E(s').
\end{align}
These additional terms push the system out of equilibrium.  When a transition has such an additional force, we say that it is ``pumped.''  Pumped systems of this form represent a general class of nonequilibrium systems, where external forces create persistent flows and entropy production.  Fig. \ref{fig:PumpedNetwork} visualizes such a thermal network, highlighting pumped transitions in red and thermal transitions in black.

\begin{figure}
\includegraphics[width=1\columnwidth]{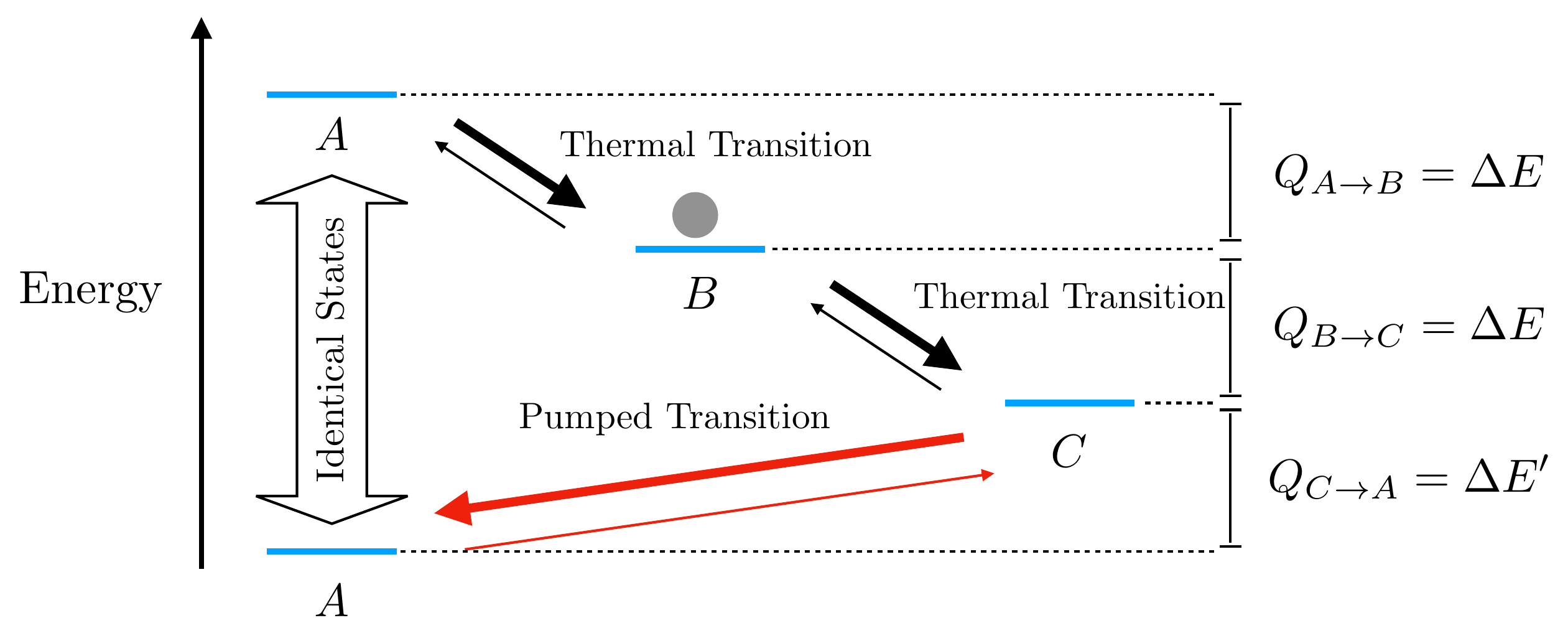}
\caption{\emph{Three-state pumped nonequilibrium system: }The energy landscape descends by a value $\Delta E$ from $A$ to $B$, then by the same amount from $B$ to $C$.  This sets the energy landscape up to a constant factor, which would make the transition from $C$ to $A$ highly unlikely in equilibrium.  The transition from $C$ to $A$ is pumped by an external force to create persistent nonequilibrium currents.  This results in the heat dissipation $\Delta E'$ in the transition from $C$ to $A$.} 
\label{fig:PumpedCycle} 
\end{figure}

\begin{figure*}
\includegraphics[width=2\columnwidth]{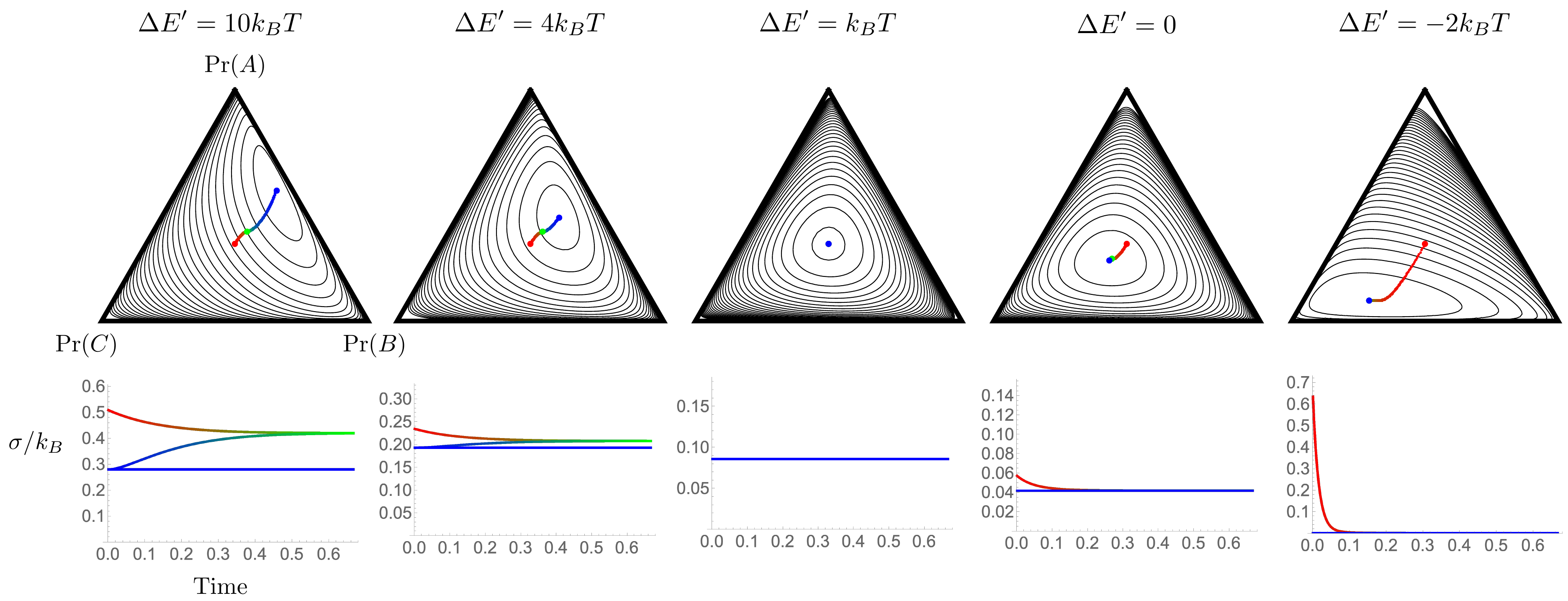}
\caption{The entropy production rate of the rate equation $R$ over the simplex of system states $A$, $B$, and $C$ for a variety settings for the pumping parameter $\Delta E'$ and $\Delta E= k_B T$.  For each we plot the contours of the entropy production evaluated at every point in the 3-simplex specifying $(\Pr(A),\Pr(B),\Pr(C))$.  Within each simplex, we highlight the uniform state $u$ (red dot), NESS $\pi$ (green dot), and MEPS $m_R$ (blue dot).  We plot how the uniform and MEPS evolve through the simplex to the NESS, and track how the entropy production changes in time the plot below each simplex.  From these, we see that MEPS and NESS are clearly distinct for some parameter settings. }
\label{fig:EntSimplex} 
\end{figure*}

\begin{figure*}
\includegraphics[width=2\columnwidth]{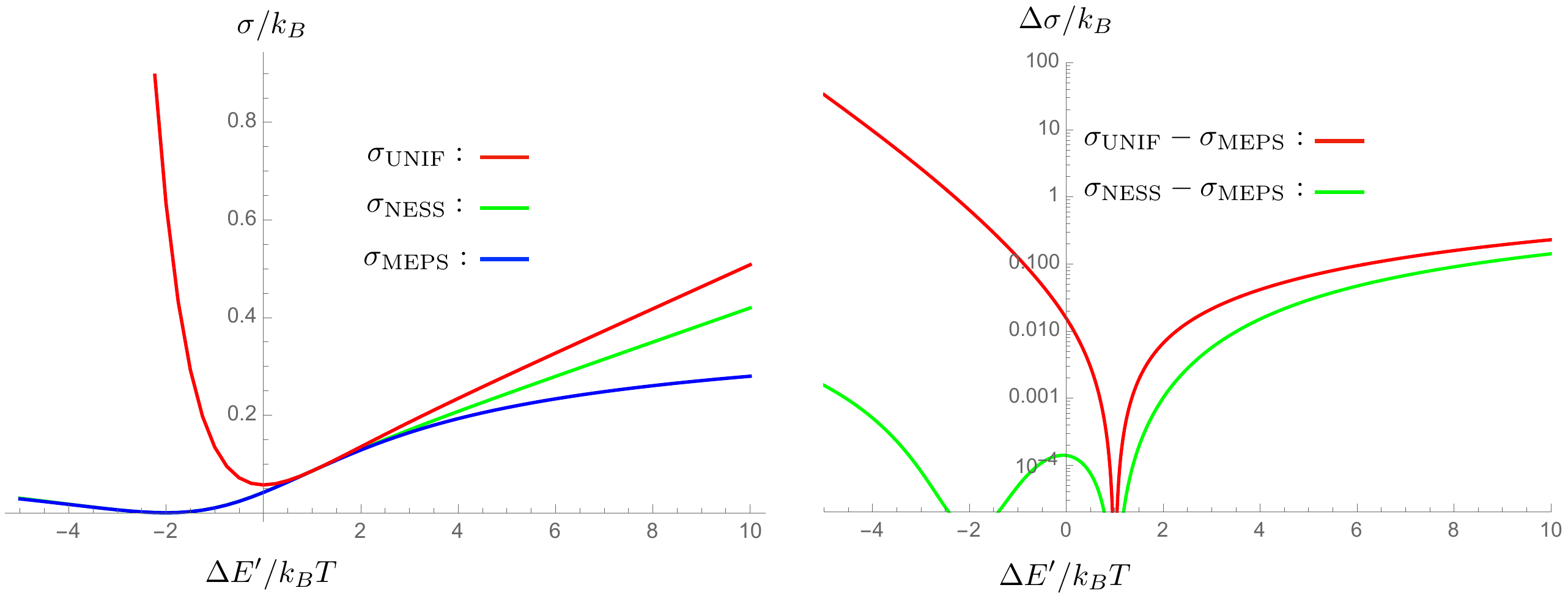}
\caption{As we change the pumping parameter $\Delta E'$, we see that the relationship between entropy production $\sigma$ of the MEPS (blue) and NESS (green) changes.  (Left) By plotting the entropy production of the MEPS, NESS, and uniform state together, we see that the MEPS achieves zero dissipation for a single energy value, where $\Delta E'=-2$.  This is thermodynamic equilibrium.  We also see that the entropy production of the uniform distribution (red) is always above the NESS and MEPS distributions for these parameters. (Right)  By plotting the additional entropy production beyond the MEPS for the NESS and uniform state, we see that the entropy production of NESS and MEPS are often close for low and negative values of $\Delta E'$, but the steady state entropy production diverges from the MEPS for higher positive values of $\Delta E'$. Additionally, there are two points where the entropy produced by each state is highly symmetric. At these points only, the NESS achieves the minimum: at thermodynamic equilibrium where $\Delta E'=-2$, and when the cycle is symmetric with regard to the states at $\Delta E'=1$.}
\label{fig:EntropyComparison} 
\end{figure*}

\subsection{A Small Physical Counterexample}

We examine the claim that the steady state minimizes the entropy production with a class of simple three-state system $\mathcal{S}=\{A,B,C\}$ that is driven through cyclic trajectories by a free energy resource, much like ATP synthase \cite{brown2019theory, leighton2024flow} or the pumped energy levels of a three-level laser system \cite{andrews2010off}. As shown in Fig. \ref{fig:PumpedCycle}, the transitions $A \leftrightarrow B$ and $B \leftrightarrow C$ are both mediated by thermal fluctuations, and probabilistic transitions are made according to decreasing energies from $A$ to $B$ to $C$.  In equilibrium, this system would collect probability mass in the low-energy state $C$, but it is also pumped from $C$ to $A$, driving the nonequilibrium cycle $A \rightarrow B \rightarrow C \rightarrow A \cdots$.  The force from $C$ to $A$ is designed such that it appears as if there is an energy change of $\Delta E'$ from $C$ to $A$, resulting in heat $Q_{C \rightarrow A}= \Delta E'$.  This apparent change in energy is the result of a force that modulates the transition from $A$ to $C$:
\begin{align}
    F_{A \rightarrow C}=-2\Delta E - \Delta E'.
\end{align}
All other forces, including $F_{C \rightarrow A}$ are zero here.  This single forced transition expedites reverse transitions when it takes positive values and restricts those same transitions when it is negative.  App. \ref{app:Three-State System} describes the mechanics of this system in greater detail.

This thermal system behaves much like a spiral staircase, where a ball tends to take a bouncing spiral down the stairs, driven by gravity when viewed from above.  We don't directly observe which floor the ball is on in the same way that the experimentalist might not directly observe the concentration of nonequilibrium resources that drive the transition from $C$ to $A$.  

In exploring the pumping parameter $\Delta E'$, we find a clear counterexample to entropy minimization.  Figs. \ref{fig:EntSimplex} and \ref{fig:EntropyComparison} show that in the simplex of a three-state nonequilibrium system, the NESS and MEPS distributions are generically distinct $m_R \neq \pi$.  In addition, Figs. \ref{fig:EntSimplex} and \ref{fig:EntropyComparison} show that the entropy production in steady state $\sigma_\text{NESS}$ is above the minimum value $\sigma_\text{MEPS}$. We also identify two cases in which the MEPS and NESS distributions collapse to the same thing: equilibrium, and a uniform cycle. In these cases, $\sigma(R,m_r,s) = \sigma(R,m_r,s')$, so this collapse is in accordance with equation \ref{eq:meps_dynamics}, which identifies such symmetric cases as simultaneously solving the condition for MEPS and NESS.

While the entropy production rate of the steady state is often well above the minimum entropy production, we also see that it is always less than or equal to the EPR of the uniform distribution.  Thus, while entropy minimization is not the general trend of thermodynamic processes, it appears that the steady state often improves upon thermodynamic efficiency relative to many other states.  We have not yet formalized this advantage, except to observe that the NESS typically has a reduced EPR when compared not only to the uniform distribution but also distributions with state probabilities randomly sampled from the Dirichlet distribution, as we will show in the next section. 

\subsection{Entropy Minimization in Dense Networks}

Next, we investigate what happens as these nonequilibrium pumped systems grow large. As we examine large statistical collections of thermodynamic systems, we introduce two measures to identify the proximity to MINEP. 
\begin{enumerate}
    \item The \emph{supraminimal EPR} of a state $p$ is the entropy production beyond the minimum $\sigma(R,p)-\sigma_\text{MEPS}$, and it is therefore non-negative. We evaluate the claim of MINEP by examining the \emph{scaled supraminimal EPR} of the steady state 
\begin{align}
    \frac{\sigma_\text{NESS}-\sigma_\text{MEPS}}{\sigma_\text{MEPS}}=\frac{\sigma_\text{NESS}}{\sigma_\text{MEPS}}-1
    ~.
\end{align} 
Since systems that are near equilibrium are guaranteed to nearly achieve the bound due to detailed balance, we use this scale factor to normalize across different non-equilibrium systems that vary in their proximity to equilibrium.
\item The \emph{supraminimal relative entropy} is Kullback-Leibler divergence between the steady state and the MEPS distribution
\begin{align}
    D_{KL}(\text{NESS}|\text{MEPS}) \equiv \sum_{s} \pi(s) \ln \frac{\pi(s)}{m_R(s)}.
\end{align}
This is the probabilistic divergence of the NESS from the MEPS distribution.
\end{enumerate}
These measures address the question of MINEP from different and complementary directions: the scaled supraminimal EPR determines how close the entropy production is to the theoretical minimum, while the supraminimal relative entropy determines how close the distribution is to the MEPS.  In what follows, we will see that these two measures largely agree on when MINEP is reached.

In all simulations, an equilibrium system was first established by choosing $N$ state energies $E_s$ and $\frac{1}{2}(N^2-N)$ barriers $E^{\text{barrier}}_{s\leftrightarrow s'} = \text{max}(E_s,E_{s'})+ E_{s\leftrightarrow s'} $ with $E_{s\leftrightarrow s'} \in [0, 1]\cdot k_B T$ and $E_s \in [-1, 1]\cdot k_B T$, both sampled uniformly.  The resulting thermodynamic equilibrium state has variance, but that variance is bounded $e^{-2} \leq \frac{p_\text{eq}(s)}{p_\text{eq}(s')} \leq e^{2}$. Then, nonequilibrium dynamics were added to the system in accordance with equation \ref{eq:Pumps} by choosing a percent of total transitions to pump, and sampling effective pump energies uniformly with $F_{s \rightarrow s'} \in [-\alpha, \alpha] \cdot k_B T$, with $\alpha$ being a ``pump strength" parameter that characterizes the maximum strength of a pump when compared to the thermal energy scale of the system's equilibrium dynamics.\footnote{The standard deviation of $F_{s \rightarrow s'}$   is equal to $\alpha$ times the standard deviation in $E_s$.} In our tests, this ranges from 25\% to 500\%.

\begin{figure*}[htbp]
\centering
\includegraphics[width=2\columnwidth]{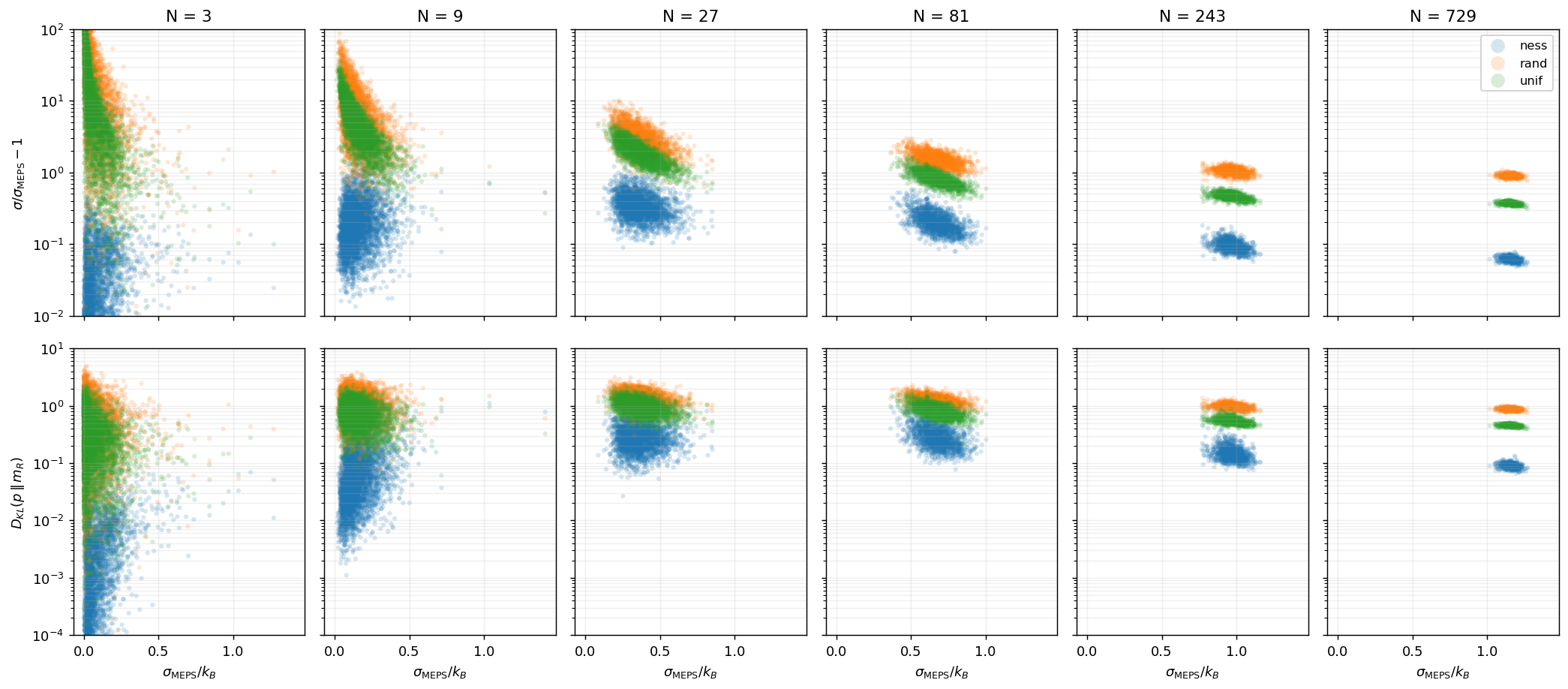}
\caption{ (TOP) Densely connected thermodynamic networks with many degrees of freedom appear to self-organize to minimize entropy production.  We compare the scaled supraminimal EPR in steady state $\sigma_\text{NESS}/\sigma_\text{MEPS}-1$ (blue) to that of uniform states $\sigma_\text{UNIF}/\sigma_\text{MEPS}-1$ (green), and randomly selected states according to the Dirichlet distribution $\sigma_\text{RAND}/\sigma_\text{MEPS}-1$ (orange).  As the number of states increases, we see that the scaled supraminimal entropy production for random and uniform states converges to approximately unity, while it is approximately an order of magnitude lower for steady state.  This indicates that steady states are generally more efficient than randomly selected distributions in the simplex.
(BOTTOM) As the number of states $N$ of a dense thermodynamic network increases, it is increasingly typical for the NESS to closely match the MEPS, as seen through the supraminimal relative entropy $D_{KL}(\text{NESS}||\text{MEPS})$, which falls an order of magnitude below the relative entropy between the MEPS and the uniform and random distributions.
These nonequilibrium systems are set up with pump strength = 400$\%$, and the percent of pumps = 80$\%$.}
\label{fig:EntComplexity}
\end{figure*}

\begin{figure*}[htbp]
\centering
\includegraphics[width=2\columnwidth]{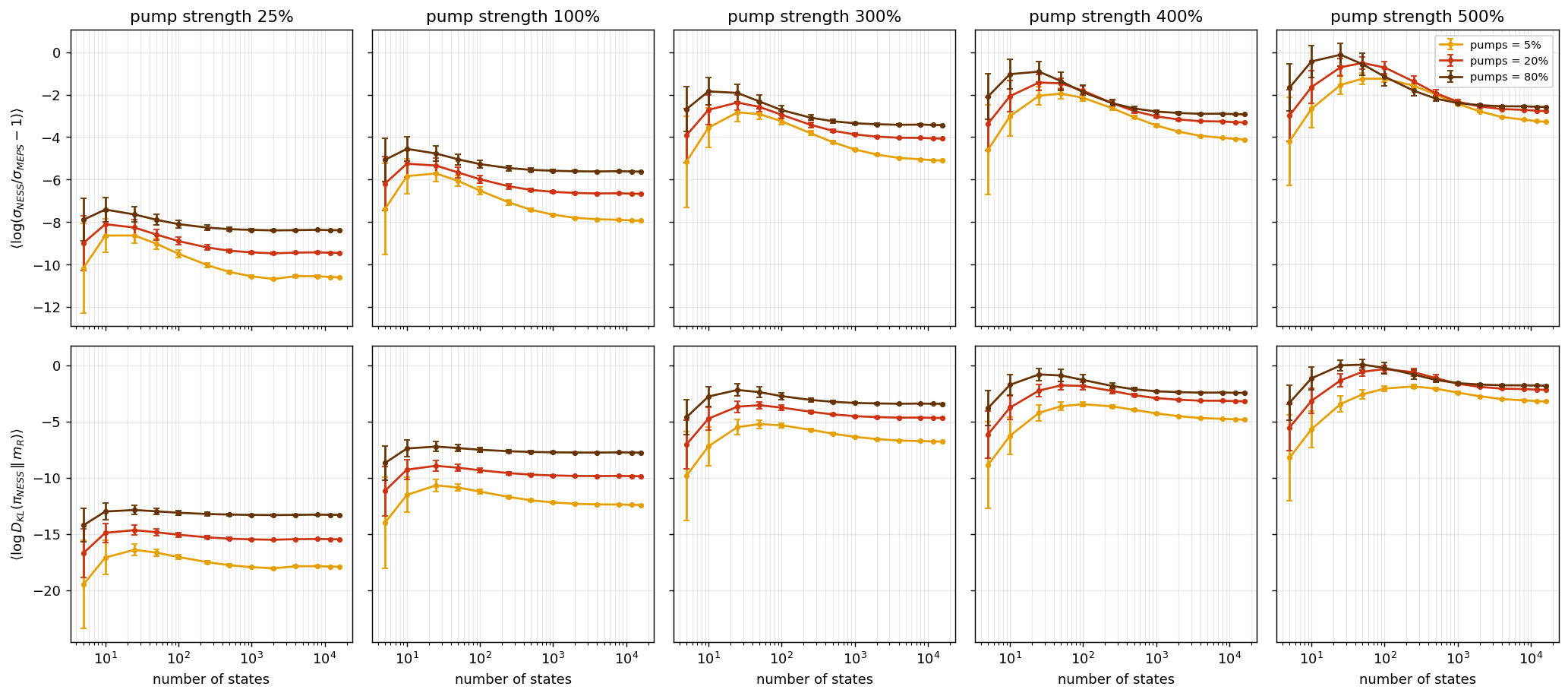}
\caption{We see that the minimum entropy production increases with the number of states (complexity) of the pumped model, especially when pumps push the system far from equilibrium.  However, we also see that the steady-state EPR scaled by the minimum EPR $\sigma_\text{NESS}/\sigma_\text{MIN}-1$ (scaled supraminimal EPR) approaches a regime in all cases where it decreases as the number of states $N$ increases. Thus, it appears that complex systems tend to minimize entropy production.  The same trend appears when we examine the supraminimal relative entropy.}
\label{fig:scaledEPRdiff}
\end{figure*}

We numerically demonstrate the typical entropic benefit of the steady state in Figs. \ref{fig:EntComplexity} and \ref{fig:scaledEPRdiff}.  Fig. \ref{fig:EntComplexity} plots scaled supraminimal EPR and supraminimal relative entropy for randomly selected non-equilibrium rate matrices with $N$ states for $N \in \{3,9,27,81,243,729\}$. Broadly, it shows clusters of steady state distributions (blue) typically have a lower EPR than uniform distributions (green) and randomly assigned distributions (orange).  However, we see differing behavior for different-sized systems. 

Starting with the simple case of $N=3$, we see large variations in both supraminimal EPR and relative entropy.  MINEP is clearly not generally satisfied here.  Many systems nearly approach the minimum entropy production and distribution nearly match the MEPS, and many others diverge from the MEPS and exceed the minimum entropy production by a significant amount.  

However, as we increase the degrees of freedom in the sampled rate equations by increasing the number of states, we see a refinement of the entropy production rate statistics.  The variance of the EPR and relative entropy decrease, and the cluster of steady states clearly separates from the uniform and randomly selected distributions.  For $N=729$, not only does the steady state typically dissipate far less entropy than these cases, but it is only marginally above the MINEP distribution ($\sim 5 \times 10^{-2}k_B$), even when $\sigma^\text{min}$ is nonzero ($\sim k_B$).  According to these samples, it appears that larger dimension thermodynamic processes nearly minimize entropy production.

In addition to showing low supraminimal entropy production of the steady state, Fig. \ref{fig:EntComplexity} reveals a special property of the steady state.  For large systems, the NESS almost always dissipates less than a randomly selected state or the uniform state.  This means that, if such a large thermodynamic network were randomly configured in a distribution, it would almost certainly relax to a more thermodynamically efficient steady distribution.  This echoes a form of adaptation observed in thermodynamic spin-glasses \cite{gold2019self}, where systems spontaneously organize to reduce the use of thermodynamic resources.  Thus, while MINEP is only ``nearly'' true, a more descriptive principle might be \emph{reduced entropy production (REDEP)}: that the steady state typically reduces the entropy production from a randomly selected initial state. 

Fig. \ref{fig:scaledEPRdiff} expands the analysis of Fig. \ref{fig:EntComplexity} by plotting the average and variance of the supraminimal EPR and relative entropy for much larger networks.  The plots show variable pump strengths and the fractions of pumped transitions.  The broad picture that emerges is that, regardless of the pumping parameters, the system departs from MINEP as the network size increases, but eventually approaches MINEP with relatively low supraminimal EPR and relative entropy with very low variance.

It is worth noting that the example systems in Figs. \ref{fig:EntComplexity} and \ref{fig:scaledEPRdiff} are densely connected in the strictest sense of the word: every edge between any two states has a nonzero value in the rate matrix.  However, we can loosen our definition of dense to include networks where edges are cut, but the number of edges is roughly proportional to $N_\text{states}^2$, meaning that the fraction of nonzero edges remains constant as we expand the size of the network.  App. \ref{app: Additional Experiments} explores such networks numerically, going down to 10 percent edge fraction.  The overall trend holds for such networks, where the average and variance of supraminimal entropy production reduces for larger networks.

While \emph{any} subset of matrices we sample will be measure-zero in the subset of possible dense matrices, we strive to show the generality of these results for dense systems in App. \ref{app: Additional Experiments}.  We also consider two new cases:
\begin{enumerate}
    \item Pump strength $F_{s \rightarrow s'}$ is strictly positive, meaning that pumps only boost probability flow.
    \item Pump strength $F_{s \rightarrow s'}$ is strictly negative, meaning that pumps only suppress probability flow.
\end{enumerate}
Once again, we conclude that this variation on the way we sample rate-matrices does not meaningfully affect the conclusions stated above.  MINEP is nearly satisfied for large dense networks.

\subsection{Counterexample}

One specific counterexample to near MINEP arises by creating a low-activity sub-component of the dynamics. Consider any of the networks sampled in the previous examples, characterized by a rate matrix $R_{s \rightarrow s'}$ over the state space $\mathcal{S}$.  We can construct the network with a separation of time scales by selecting a special state $s^* \in \mathcal{S}$, such that the transitions to and from that state are scaled by a factor $r\ll 1$.  The result is a new rate matrix $R'_{s \rightarrow s'}$ where original rates between elements of $\mathcal{S}$ are the same while and only the transitions involving $s^*$ are changed:
\begin{align*}
    R'_{s \rightarrow s'} & = R_{s \rightarrow s'} \text{ for }s,s' \in  \mathcal{S}-\{s^*\}
    \\ R'_{s^* \rightarrow s'} & = r R_{s^* \rightarrow s'} \text{ for }s' \in  \mathcal{S}-\{s^*\}
    \\ R'_{s \rightarrow s^*} & = r R_{s \rightarrow s^*} \text{ for }s \in  \mathcal{S}-\{s^*\}
\end{align*}
With diagonal entries set by $\sum_{s'} R'_{s \rightarrow s'} = 0$. Here, $r$ is the time-scale separation factor that highlights the difference between the dynamics on the larger state subspace $\mathcal{S}-\{s^*\}$ and the state $s^*$.

\begin{figure*}[htbp]
\centering
\includegraphics[width=2\columnwidth, trim={0 0 0 60}, clip]{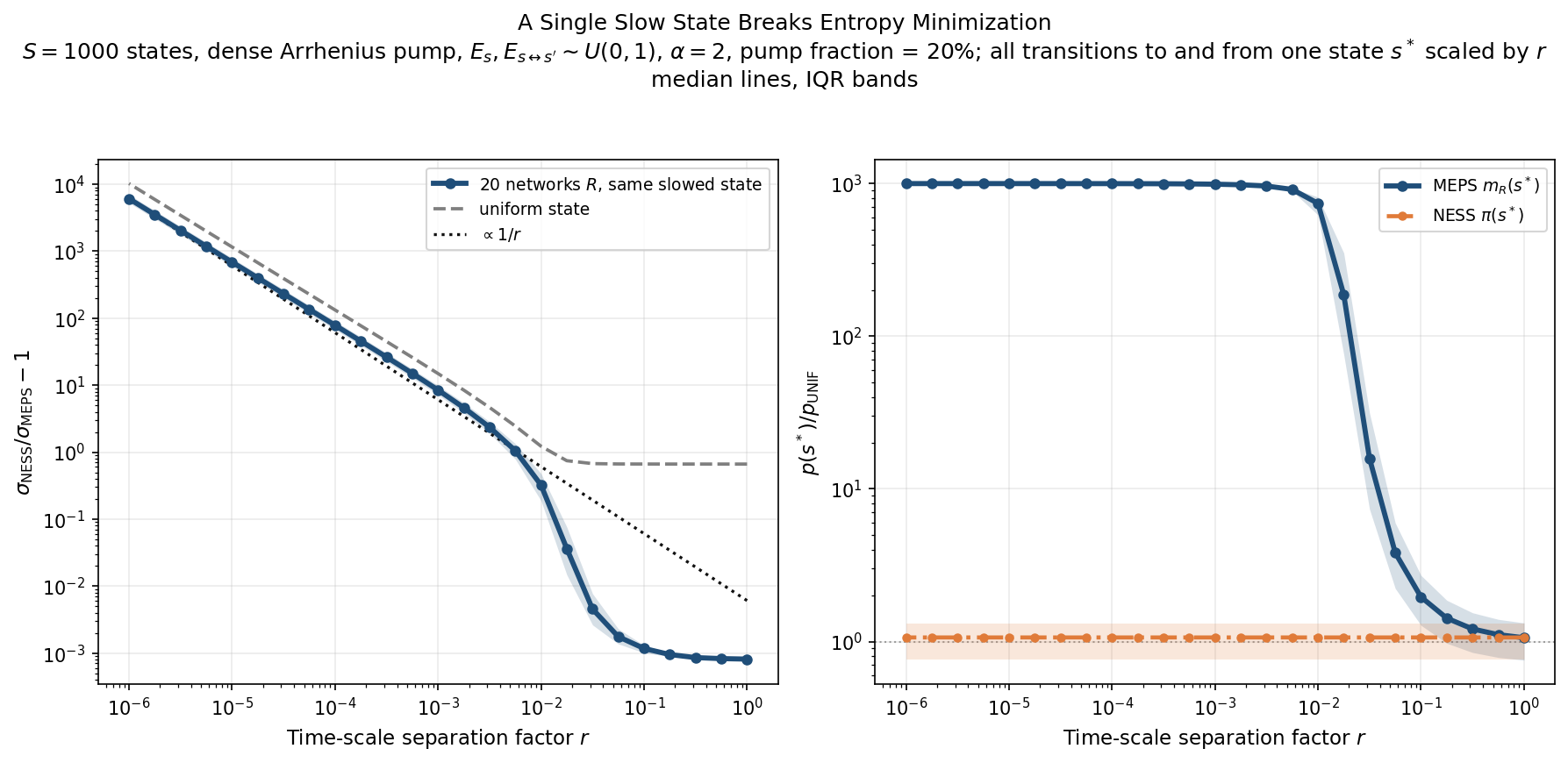}
\caption{\emph{A single low-activity state breaks entropy minimization:}  Starting from dense pumped Arrhenius networks with $S=1000$, pump fraction $20$ percent, and pump strength $200$ percent, we select one state $s^*$ and scale every transition into and out of it by the time-scale separation factor $r$.  (Left) the scaled supraminimal EPR $\sigma_\text{NESS}/\sigma_\text{MEPS}-1$, together with the same measure evaluated at the uniform state rather than the steady state (gray, dashed). Below $r \approx 10^{-2}$, the supraminimal EPR grows without bound, tracking the $1/r$ guide (dotted). (Right) the probability that the MEPS and the NESS assign to $s^*$, reported relative to the uniform value $1/S$. The MEPS localizes onto $s^*$, where almost no entropy is produced, saturating only once it holds essentially all of the probability while the steady-state occupation remains nearly the same across the entire sweep. (lines display the median over $20$ networks, with shaded regions for the interquartile range.)}
\label{fig:Counterexample}
\end{figure*}

In the limit as $r \Rightarrow 0$, the timescale of $s^*$ is so slow that there is effectively no entropy production from that state.  That means that MEPS state will approach a distribution that is peaked around the $s^*$ state and produces nearly zero entropy.  By contrast, the steady state of the whole system $\mathcal{S}$ will maintain significant probability outside of the low activity state $s^*$, and therefore produce nonzero entropy production, as shown in Fig. \ref{fig:Counterexample}.  This suggests that separation of time-scales can lead to strong MINEP violation.

This case does not arise randomly in our investigation of dense thermodynamic networks because of the statistics of sampling edges.  Low-rate edges are low-probability fluctuations, and it is unlikely that a single state, when it is connected to $\mathcal{O}(N_\text{states})$ other nodes in a large network, will have all of its randomly sampled connections realize low rates.  However, for sparser networks, it may be that the fluctuations in edge weights can lead to separations in time scales that drive away from MINEP.  We examine this briefly in the next section.

Thus, we see that strong violations of MINEP arise in time-scale separated systems where symmetry-breaking in the network is significant.  Time-scale separation involves subcomponents with less probability flow, so the dynamics are highly asymmetric under state exchange.  This appears to propagate into a violation of the entropy-production symmetry condition expressed Eq. \ref{eq:Ent_Symmetry}, and, therefore, into a violation of the equality between NESS and MEPS.  While the conditions on dynamics that lead to entropy-production symmetry are still opaque, this example of strong MINEP violation indicates a connection between dynamical asymmetries and inefficiency.

\subsection{Entropy Minimization in Sparse Networks}

\begin{figure*}[htbp]
\centering
\includegraphics[width=2\columnwidth]{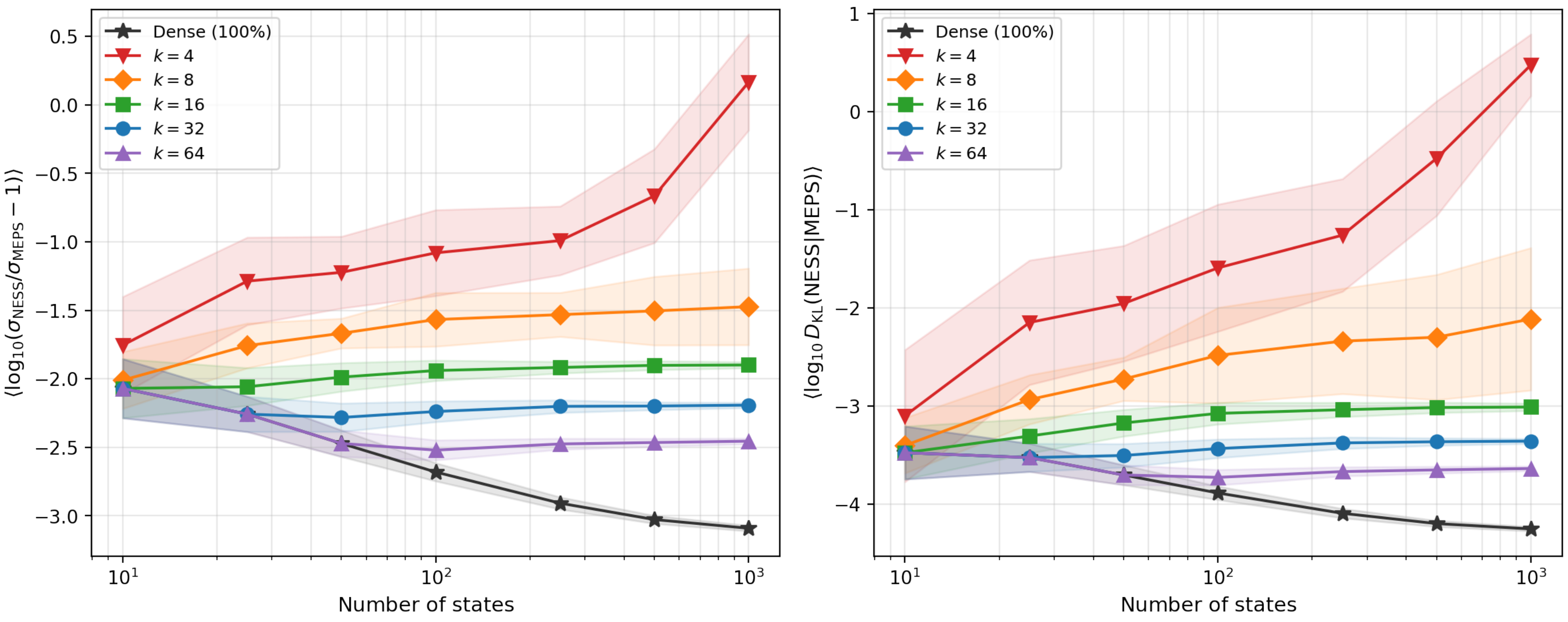}
\caption{\emph{MINEP is strongly violated for large sparse networks:} We plot the average log of scaled supraminimal EPR and supraminimal relative entropy networks with fixed average degree (edges-per-node).  For a low number of states, all the plots roughly overlap, but as the number of states increases, the denser networks decrease their supraminimal values, more closely approaching the MINEP regime, while sparser networks increase, diverging from MINEP.  (100 trials per data point, shaded region represents the standard deviation of the $log$ value, pump fraction is 20 percent, and pump strength is 200 percent).  }
\label{fig:Sparse_Dissipation}
\end{figure*}

To put the claim that large dense networks nearly minimize entropy production into relief, we consider the alternative sparse networks.  Rather than the number of edges scaling as $N_\text{states}^2$, we consider a \emph{fixed} average degree $k$, meaning that the number of edges increases with the number of states $N_\text{edges} = \mathcal{O}(N_\text{states})$.  Fig. \ref{fig:Sparse_Dissipation} interpolates between dense networks and sparse ones by considering ever-smaller degrees.  Large-degree networks dissipate less, though their supraminimal entropy production appears to level off with increasing number of states.  However, for small degree $k=4$, we see that increasing state-number actually increases the supraminimal entropy production by orders of magnitude, inverting the relationship between size and efficiency that we saw for dense networks.

\section{Speculation and Future Work}

Attempts to explain emergent organization in nonequilibrium systems, rooting the science of life and intelligence in fundamental physics, have often focused on entropy and thermodynamic resources \cite{prigogine1946biology, Fris10a, harte2008maximum, boyd2022thermodynamic, horowitz2017spontaneous, harte2011maximum}.  The claim that thermodynamic resource optimization is connected to learning and intelligence is gaining credence \cite{ Fris10a, stern2024physical}.  Specifically, minimizing dissipation is formally equivalent to learning in a variety of contexts: supraminimal entropy production equals a model–environment loss in the form of relative entropy \cite{kolchinsky2017dependence, boyd2022thermodynamic}, retained nonpredictive information carries an irreducible thermodynamic cost \cite{Stil12a}, and the leading generative models are trained by minimizing trajectory entropy production \cite{sohl2015deep, ikeda2025speed}.  However, it leaves open the question of whether MINEP is indeed typical.  In essence, there are strong physical links between entropy production minimization and learning.  Given this, should we expect matter to spontaneously self-organize into living, learning systems?  A key step in determining the answer to this question is to understand the laws governing entropy production and the typicality of its minimization.

Stochastic thermodynamic continuous-time Markov chains are an ideal framework for addressing the general relationship between entropy production and organization, because of the direct relationship between heat and dynamical transitions.  Within this framework, we see that entropy production is not minimized in general, meaning that we should not conclude that spontaneous learning is the universal law nonequilibrium organization.  However, surprisingly, we see that a variety of systems nearly achieve minimum entropy production in their steady state.  Specifically, the larger a dense thermodynamic network is, the more typical nearly minimum entropy production appears to be.  If you were to randomly select a configuration of the system, the nonequilibrium system will almost surely evolve to a steady state that dissipates an order of magnitude less entropy. This reflects a type of learning, where the system adapts to the nonequilibrium source that drives it \cite{gold2019self}. The fact that larger systems express this fact more strongly suggests that complexity precedes MINEP and learning.  However, there is still much more work to do to make these connections concrete.

Here, we enumerate a short list of possible follow-on projects to clarify the picture of MINEP and spontaneous organization:
\begin{enumerate}
\item \emph{Development of a complexity measure for CTMCs and comparison to EPR:}  Complexity is a deep and nuanced field with popular measures ranging from Kolmogorov-Chaitin complexity  \cite{Chai66, Kolm65} to statistical complexity \cite{Crut12a}, and beyond \cite{lohr2012predictive, gu2012quantum}. Each measure provides the answer to a different question. If a complexity measure can be determined that is appropriately suited for CTMCs, it would be revealing to compare the complexity to the scaled supraminimal entropy production.  We hypothesize that more complex nonequilibrium systems would be more thermodynamically efficient.

\item \emph{Exploring new sampling methods for CTMCs:}  An admittedly limiting factor in our results is the fact that our algorithms for randomly sampling CTMCs indirectly determine the statistics of entropy production.  While our pumped models are rooted in pumped stochastic systems and chemical reaction networks, there are many other methods one might choose, which might reflect ``typical'' systems, depending on the context.

\item \emph{Random matrix theory of rate equation entropy production:}  We anticipate that systematic predictions for EPR of large random nonequilibrium systems may be accessible through random matrix theory \cite{busiello2017entropy}, which can predict the statistical behavior of measures like the singular values that arise from random matrices.  Specifically, we hope to predict the behavior of the MEPS $m_R$ and NESS $\pi$.

\item \emph{Decomposing nonequilibrium systems into agent and environment:} If a living agent is learning about its environment in a perception-action loop \cite{fiderer2025work}, both are nonequilibrium systems, operating in unison.  They produce entropy both individually and as a whole \cite{Boyd17a}.  Comparing the information flow between agent and environment to the entropy production of both may reveal fundamental relationships between learning and MINEP.

\item \emph{Refined Definitions of MINEP and MAXEP:} We note that the definitions of MAXEP and MINEP presented here are not universally agreed upon.  In particular, rather than extremizing entropy production rate over the continuous space of probability distributions $p$, one may consider multistable systems that exhibit several distinct metastable ``steady'' states or ``attractors.'' In such cases, Refs. \cite{cook2020thermodynamics,endres2017entropy} report that among the admissible steady states, the one with the largest EPR is statistically favored, claiming MAXEP as a state-selection principle.  By contrast, Ref. \cite{falasco2025macroscopic} examines similar multistable systems and finds that attractors with \emph{minimal} EPR are the most stable, effectively establishing another form of MINEP that applies far from equilibrium, grounded in the multiplicity of steady states. These divergent interpretations highlight that even within a similar formal setting, the operational meaning of ``maximization'' or ``minimization'' remains context-dependent and unresolved, indicating a promising direction for further clarification.  
\end{enumerate}

\section{Conclusion}

The results presented here are an entrée to a new line of inquiry.  We provide a comparison to past results, showing the pitfalls of classic MAXEP and MINEP in stochastic thermodynamics.  We provide the tools for numerical calculation of the MEPS and comparison to the NESS, comparing the minimum entropy production to that in steady state.  We then show a simple example that illustrates that MINEP is not satisfied in general, but then observe that it is nearly typical for large interconnected nonequilibrium systems.

In answer to our motivating question, of whether nature ``abhors'' high entropy, we take a moderate stance.  Just as rivers sometimes meander slowly to the sea and sometimes crash towards equilibrium with sudden force, nonequilibrium systems do not obey a universal MAXEP or MINEP principle. At first blush, this is a negative result, that there is no correlation.  However, our numerical results indicate that, while CTMC dynamics do not globally minimize EPR, they do appear to typically lower it in large dense thermodynamic networks, which approach MINEP.  This effect is especially pronounced in comparison to sparse networks and those with separation of timescales, which diverge from MINEP.  These systems strongly break the MINEP entropy production symmetry condition shown in Eq. \ref{eq:Ent_Symmetry}. This opens up avenues to discuss a ``soft" form of MINEP (``SMINEP,'' perhaps). These results potentially point to a more nuanced optimization process in which EPR is only one of several contributions that a nonequilibrium system optimizes for. This emerging nuance presents us with new opportunities for discovery.

\section{Acknowledgments}

The authors thank Gavin Crooks, James Crutchfield, Vidyesh Rao, and Daniel Selsam for helpful conversations and Carlos Floyd for providing useful background resources and Gavin Crooks for introducing the low-activity counterexample. ABB acknowledges support from the Templeton World Charity Foundation Power of Information fellowships TWCF0337 and TWCF0560. This material is based on work supported by, or in part by, 
the Art and Science Laboratory and U.S. Army Research Laboratory and U.S. Army Research Office under Grant No. W911NF-21-1-0048. The python code used to generate Figs. \ref{fig:EntComplexity} and \ref{fig:scaledEPRdiff} is publicly available online \cite{kray2025}.

\clearpage
\onecolumngrid
\appendix

\section{Minimum Entropy Production}
\label{app:Minimum Entropy Production}

We apply the method of Lagrange multipliers to the entropy production
\begin{align}
    \sigma(R,p) \equiv \sum_{s,s'}p(s)R_{s \rightarrow s'} \ln \frac{p(s) R_{s \rightarrow s'}}{p(s') R_{s' \rightarrow s}}
\end{align}
by evaluating the partial derivative 
\begin{align}
\partial_{p(z)}\sigma(R,p) & = \sum_{s,s'}\left(\delta_{s,z}R_{s \rightarrow s'} \ln \frac{p(s) R_{s \rightarrow s'}}{p(s') R_{s' \rightarrow s}} + \frac{p(s)R_{s \rightarrow s'}}{p(s)R_{s \rightarrow s'}}\delta_{z,s}R_{s \rightarrow s'}-\frac{p(s) R_{s \rightarrow s'}}{p(s') R_{s' \rightarrow s}} \delta_{z,s'} R_{s' \rightarrow s}\right)
\\& = \sum_{s'}R_{z \rightarrow s'} \ln \frac{p(z) R_{z \rightarrow s'}}{p(s') R_{s' \rightarrow z}} + \sum_{s'}R_{z \rightarrow s'}-\sum_{s} \frac{p(s) R_{s \rightarrow z}}{p(z) } 
\\ & = \sigma(R,p,z)- \frac{\partial_t p(z)}{p(z)},
\end{align}
using the fact that $\sum_{s}p(s)R_{s \rightarrow z}=\partial_t p(z)$, $\sum_{s'}R_{z \rightarrow s'}=0$, and defining the state-wise entropy production
\begin{align}
    \sigma(R,p,s) \equiv \sum_{s'}R_{s \rightarrow s'} \ln \frac{p(s) R_{s \rightarrow s'}}{p(s') R_{s' \rightarrow s}}.
\end{align}
When we average the state-wise entropy production, we obtain the average entropy production
\begin{align}
\sum_{s}p(s)\sigma(R,p,s)=\sigma(R,p).
\end{align} 

%

Since the only constraint on $p$ is normalization, which can be framed as 
\begin{align}
    g(p) \equiv \sum_s p(s)=1,
\end{align}
we arrive at the minimum entropy production state $m_R$ by solving
\begin{align}
    \partial_{p(z)} \sigma(R,p)= \lambda \partial_{p(z)}g(p),
\end{align}
where $\lambda$ is the Lagrange multiplier.  Applying the fact that $\partial_{p(z)}g(p)=1$, we have the equation
\begin{align}
    \sigma(R,p,z)- \frac{\partial_t p(z)}{p(z)}= \lambda.
\end{align}
Averaging both sides over the probability, we find that the Lagrange multiplier is the average entropy production rate $\lambda=\sigma(R,p)$, and we see that the MINEP distribution $m_R$ satisfies the equality
\begin{align}
    \partial_t \ln m_R(z)=\frac{\partial_{t}m_R(z)}{m_R(z)}=\sigma(R,m_R,z)-\sigma(R,m_R).
\end{align}
The MINEP state evolves towards states that produce more entropy than average and away from those that produce less.  The steady-state is only the MINEP state if the entropy production from each state is equal.

\section{Three-State System}
\label{app:Three-State System}

We consider states with energies $E_A=E_B+\Delta E$ and $E_B=E_C + \Delta E$ in contact with a thermal energy bath at temperature $T$.  This thermal bath facilitates transitions two $A \leftrightarrow B$ and $B \leftrightarrow C$ according Arrhenius rate equations
\begin{align*}
    R_{A \rightarrow B} & =K e^{(E_A-E_{A\leftrightarrow B})/k_B T}
    \\ R_{B \rightarrow A} & =Ke^{(E_B-E_{A\leftrightarrow B})/k_B T}
    \\ R_{B \rightarrow C} & =K e^{(E_B-E_{B\leftrightarrow C})/k_B T}
    \\ R_{B \rightarrow A} & =K e^{(E_C-E_{B\leftrightarrow C})/k_B T},
\end{align*}
where the pre-exponential factor $K$ is a constant, $E_{A \leftrightarrow B}$ is the height of the energy barrier between states $A$ and $B$, and $E_{B \leftrightarrow C}$ is the height of the energy barrier between states $B$ and $C$.  However, the transition between $A$ and $C$ is not directly facilitated by thermal fluctuations in this energy landscape.  It is driven by a nonequilibrium work resource like ATP.  This driven reaction behaves as if there is an energy drop from $C$ to $A$ of $\Delta E'$ with an energy barrier $E'_{C \leftrightarrow A}$ such that the rates are
\begin{align}
    R_{C \rightarrow A} & = K e^{(E_C-E'_{C \leftrightarrow A})/k_B T}
    \\ R_{A \rightarrow C}& = K e^{(E_C-\Delta E'-E'_{C \leftrightarrow A})/k_B T}.
\end{align}
Using the equation for rates in terms of forces and $R_{s \rightarrow s'}=K e^{\frac{F_{s \rightarrow s'}}{k_BT}} e^{\frac{E(s)-E^\text{barrier}_{s \leftrightarrow s'}}{k_B T}}$

To further simplify, we choose $E_{A \leftrightarrow B}=E_\text{barrier}$, $E_{B \leftrightarrow C}=E_\text{barrier}-\Delta E$, and $E'_{C \leftrightarrow A}=E_\text{barrier}-2\Delta E$, so that the rate of transitioning forward in the cycle $A \rightarrow B \rightarrow C \rightarrow A$ is always the same
\begin{align}
R_{A \rightarrow B}=R_{B \rightarrow C} = R_{C \rightarrow A}=Ae^{E_A-E_\text{barrier}/k_B T},
\end{align}
and the reverse transition rates are exponentially damped
\begin{align}
R_{B \rightarrow A}& =R_{A\rightarrow B}e^{-\Delta E/k_B T}
\\ R_{C \rightarrow B}& =R_{B\rightarrow C}e^{-\Delta E/k_B T}
\\R_{A \rightarrow C}& =R_{C\rightarrow A}e^{-\Delta E'/k_B T}.
\end{align}

We can interpret this system as a pumped system with the rate equations determined via the relation
\begin{align}
    R_{s \rightarrow s'}=K e^{\frac{F_{s \rightarrow s'}}{k_BT}} e^{\frac{E(s)-E^\text{barrier}_{s \leftrightarrow s'}}{k_B T}}.  
\end{align} 
The pumping is restricted to the $A \leftrightarrow C$ transition, with a force $F_{A \rightarrow C}$ that modulates the reverse transition, and $F_{C \rightarrow A}=0$ .  Specifically, the apparent energy change is related to the force via the relation
\begin{align}
    F_{A \rightarrow C}= -2\Delta E-\Delta E'.
\end{align}

\section{Additional Experiments}
\label{app: Additional Experiments}

Here we report numerical experiments that vary the topology and the rate-sampling procedure of the main text.  The first family is the pumped Arrhenius network of Eq.~\ref{eq:Pumps}, sampled as in Fig.~\ref{fig:scaledEPRdiff} except that energies and barriers are drawn as $E_s, E_{s \leftrightarrow s'} \sim U(0,1)\cdot k_B T$, and every pump acts in the same direction with magnitude up to $2 k_B T$ rather than being two-sided, with $20\%$ of transitions pumped.  These dense networks are then thinned to a target edge density $f$, the retained fraction of the $S(S-1)$ possible transitions, in one of two ways.  \emph{Erdos--R\'enyi} sparsification starts with a fully connected network keeps each transition independently with probability $f$, giving average degree $k = f(S-1)$.  \emph{Small-world} sparsification instead starts with a ring lattice in which each state is joined to its $k = f(S-1)$ nearest neighbors and rewires each edge with probability $\beta_\text{SW}$.  In both cases surviving transitions inherit their rates from a same sized dense network. Since $f$ is held fixed as $S$ grows, $N_\text{edges} = \mathcal{O}(N_\text{states}^2)$ so these remain "dense".

Either sparsification can leave the chain reducible, in which case its strongly connected components are linked into a single cycle by restoring one randomly chosen transition pair between components, assigned the median rate in both directions. This repair restores ergodicity with minimal additional driving.  This affects only the small systems: even at at $f = 10\%$ no trial with $S \geq 100$ required repair.

Each figure plots the same two measures against system size $S$: the scaled supraminimal EPR $\sigma_{NESS}/\sigma_\text{MEPS} - 1$ and the supraminimal relative entropy $D_{KL}(\text{NESS}\|\text{MEPS})$. We also show the scaled entropy production rate evaluated against the uniform state (gray, dashed) rather than the NESS.  Lines are medians over trials, from $500$ trials at the smallest sizes down to $5$ at $S = 2000$, the bands represent the interquartile range. Large systems are very low variance but have a high computational cost, which is why we sampled relatively fewer of them.

\begin{figure*}
\centering
\includegraphics[width=\columnwidth, trim={0 20 0 80}, clip]{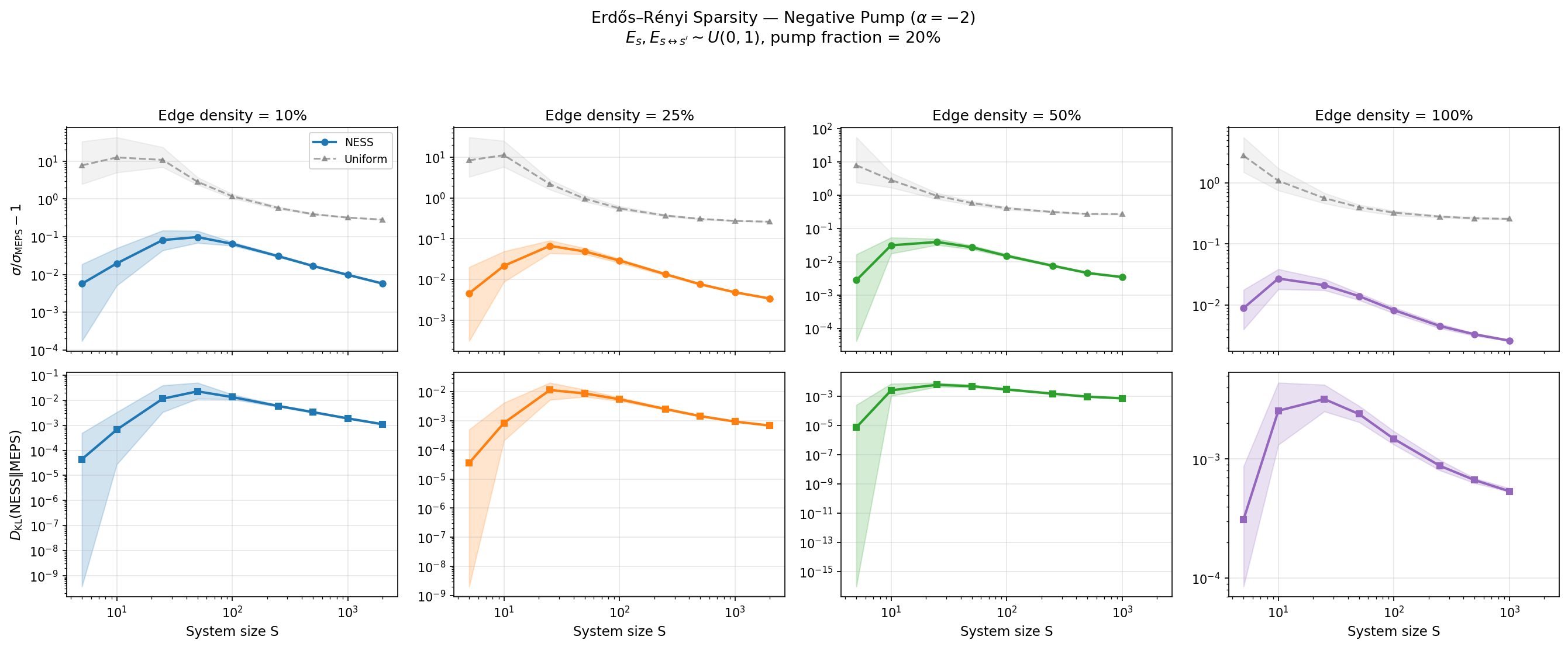}\\ 
\includegraphics[width=\columnwidth, trim={0 0 0 80}, clip]{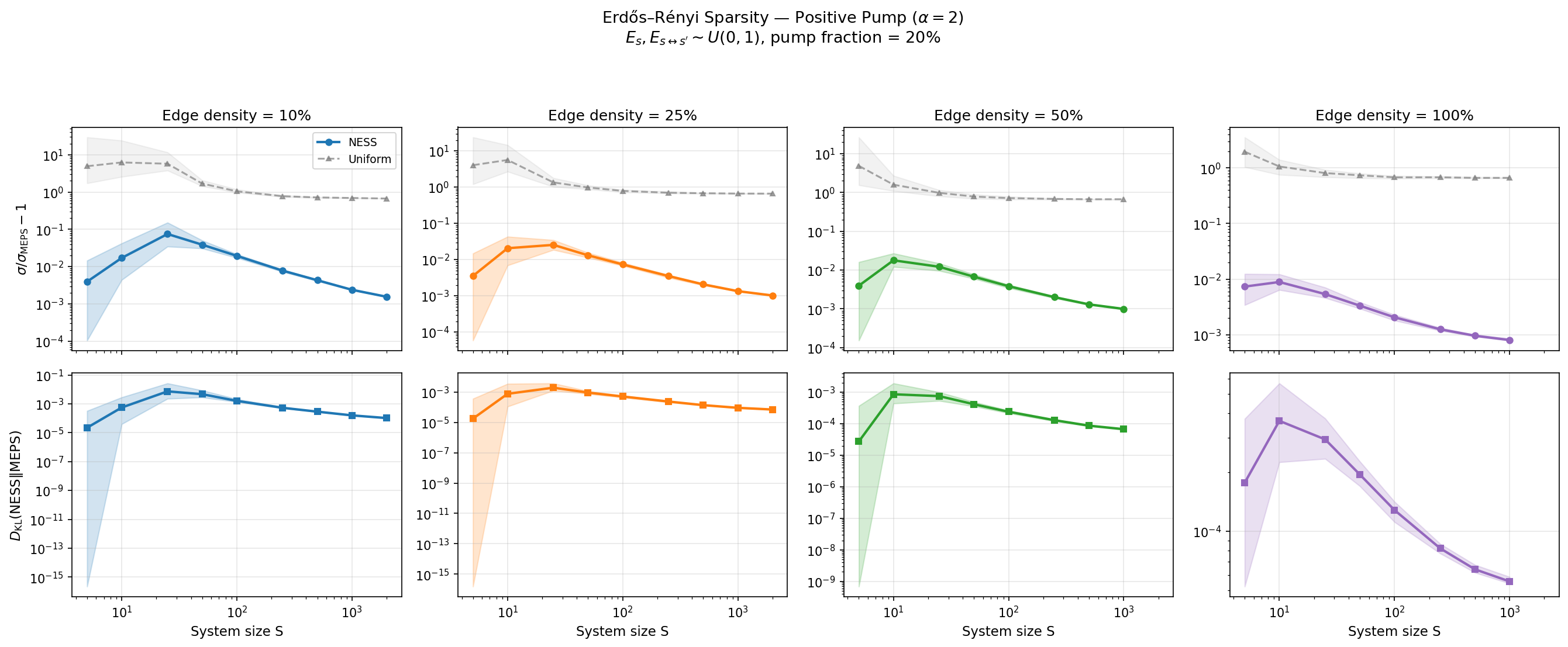}
\caption{\emph{Erdos--R\'enyi sparsification:} pumped Arrhenius networks at edge densities $f = 10\%$, $25\%$, $50\%$, $100\%$ (columns), with pumps that enhance (top, $F_{s \rightarrow s'} \in [0,2]$) or suppress (bottom, $F_{s \rightarrow s'} \in [-2,0]$) the transitions they act on. Pump fraction is $20\%$ for all cases.  Both measures fall with system size at every density and for either pump direction.}
\label{fig:appendix_er}
\end{figure*}

\begin{figure*}
\centering
\includegraphics[width=\columnwidth, trim={0 20 0 80}, clip]{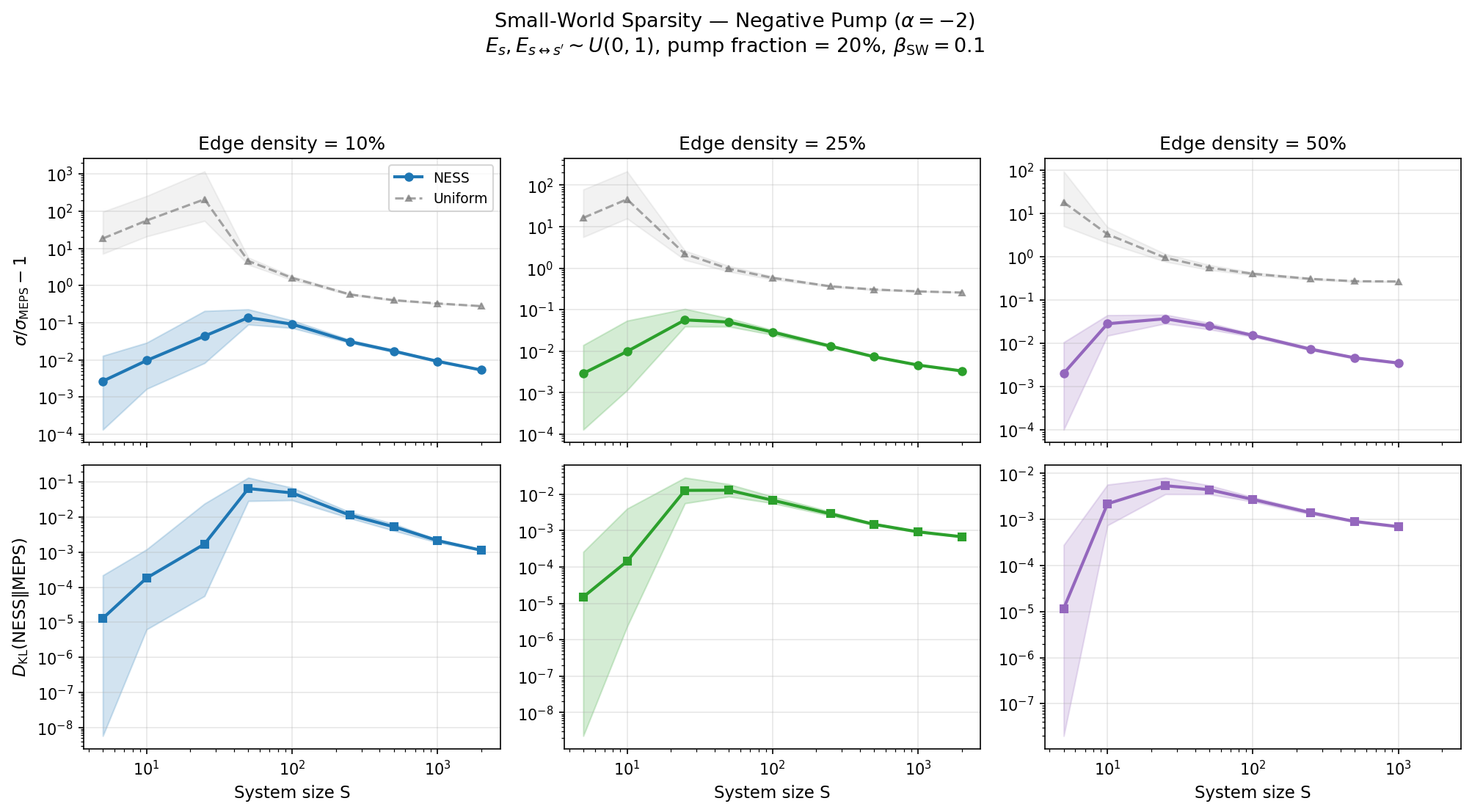}\\
\includegraphics[width=\columnwidth, trim={0 0 0 80}, clip]{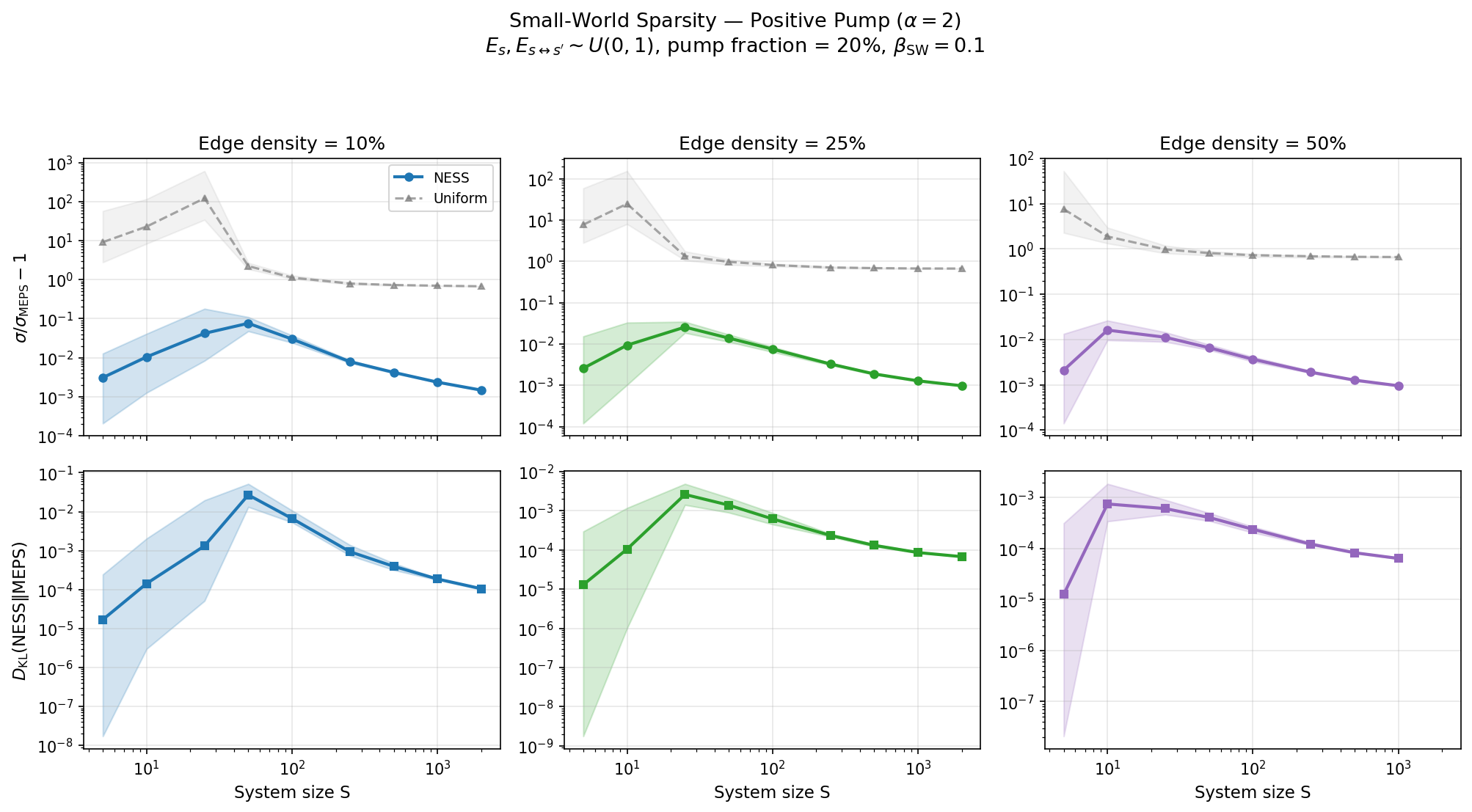}
\caption{\emph{Small-world sparsification:} as Fig.~\ref{fig:appendix_er}, with rewiring ($\beta_\text{SW} = 0.1$) in place of Erdos--R\'enyi edge deletion at matched edge density. Pumps only enhance (top, $F_{s \rightarrow s'} \in [0,2]$) or suppress (bottom, $F_{s \rightarrow s'} \in [-2,0]$) the transitions they act on. Pump fraction is $20\%$ for all cases. The clustering and short path lengths that a small-world network provide leave the trend unchanged.}
\end{figure*}

A second family samples rate equations by assigning locality to the states. The $S$ states are arranged on a ring and indexed modulo $S$, so the lattice closes on itself and the separation between two states is the circular distance $d(s,s') = \min(|s-s'|, S - |s-s'|)$; this yields states $0$ and $S-1$ as nearest neighbors, $d(0,S-1) = 1$. No two states are farther apart than $\lfloor S/2 \rfloor$.  Every pair satisfying $d(s,s') \leq d_\text{max}$ is connected, so each state has $2 d_\text{max}$ neighbors. This sets the edge density to $f = 2 d_\text{max}/(S-1)$, with $d_\text{max} = \lfloor S/2 \rfloor$ giving a fully connected ring.  Rates are drawn per state and per separation: state $s$ hops clockwise at rate $R_{s \rightarrow s+d} = e^{\mathcal{N}(0,1)}$ and counter-clockwise at rate $R_{s \rightarrow s-d} = R_{s \rightarrow s+d} \cdot U(0,1.5)$ independently for every $s$ and every $d$.  Each state thus carries a bias between the two directions around the ring; since $U(0,1.5)$ lies below unity two-thirds of the time, a net current circulates. This is the source of the nonequilibrium driving. The driving assigns each transition a heat $Q_{s \rightarrow s'} = k_B T \ln (R_{s \rightarrow s'}/R_{s' \rightarrow s})$ by local detailed balance.  The small-world variant rewires each edge with probability $\beta_\text{SW} = 0.1$, the rewired edge carrying its rates to its new endpoint so that the edge count and the distribution of rates are preserved.

\begin{figure*}
\centering
\includegraphics[width=\columnwidth, trim={0 20 0 80}, clip]{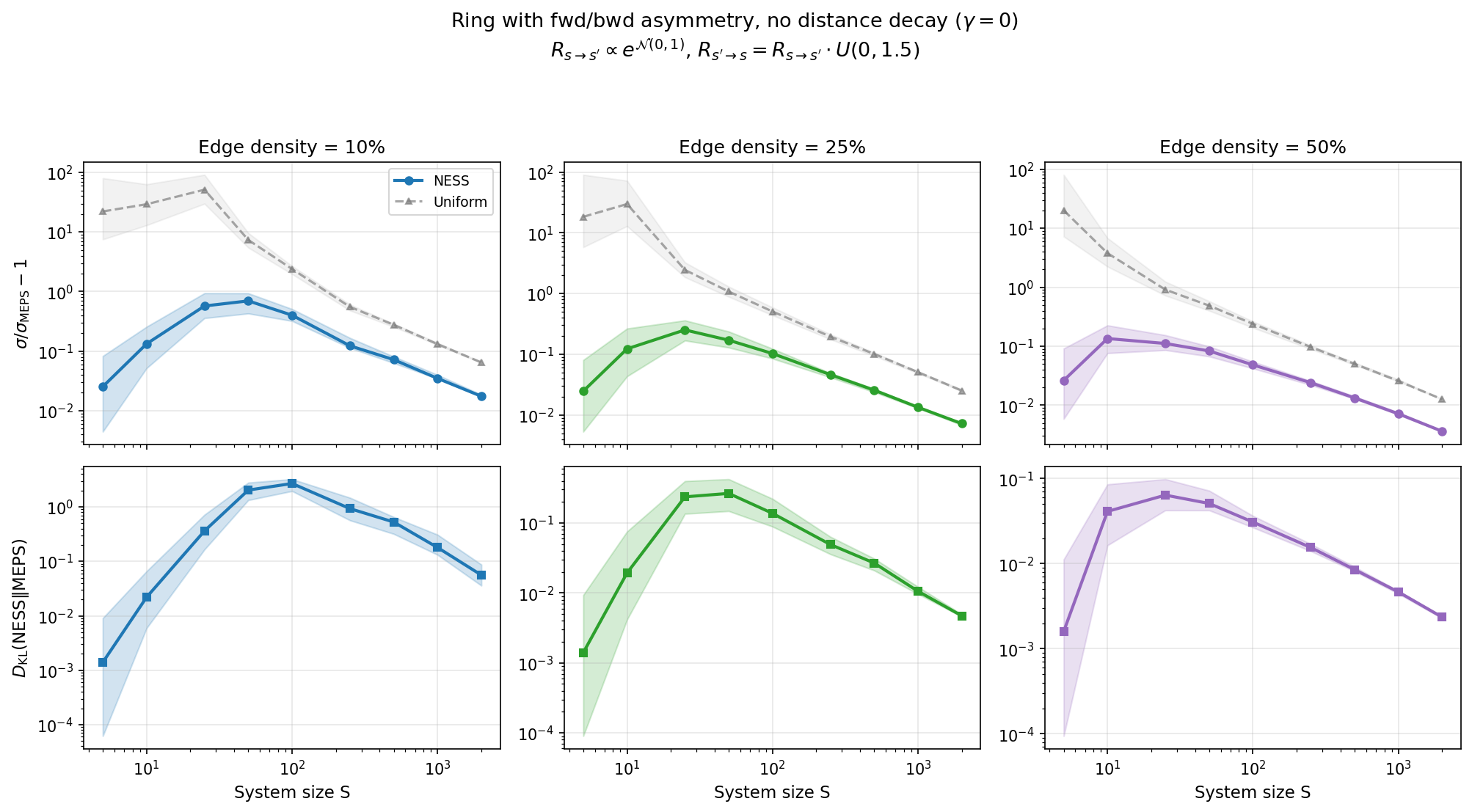}\\
\includegraphics[width=\columnwidth, trim={0 0 0 80}, clip]{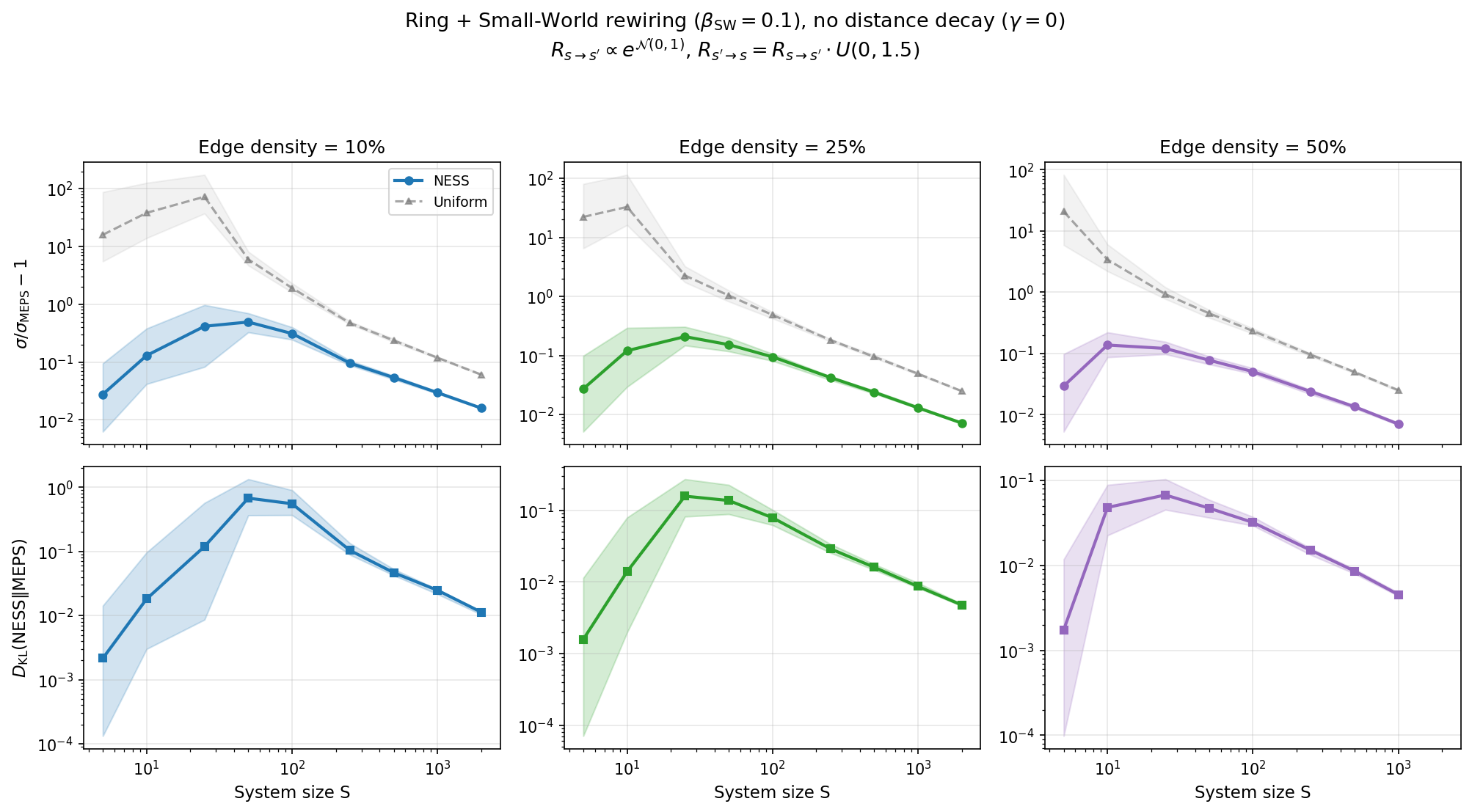}
\caption{\emph{Rings:} rings at three edge densities (columns) before (top) and after (bottom) small-world rewiring at fixed edge count.  After a transient at small $S$, both measures decrease with system size.}
\end{figure*}


\twocolumngrid
\bibliography{chaos}

\end{document}